\pdfoutput=1

\documentclass[11pt]{article}

\usepackage[final]{acl}

\usepackage{times}
\usepackage{latexsym}

\usepackage[T1]{fontenc}

\usepackage[utf8]{inputenc}

\usepackage{microtype}

\usepackage{natbib}
\usepackage{amsmath,amssymb,amsfonts}
\usepackage{algorithmic}
\usepackage{graphicx}
\usepackage{textcomp}
\usepackage{xcolor}
\usepackage{subfigure}
\usepackage{xspace}
\usepackage{mathtools}
\usepackage{enumitem}
\usepackage{multirow}
\usepackage{adjustbox}
\usepackage{tikz}
\usepackage{siunitx}
\usepackage{caption}
\usepackage{verbatim}
\usepackage{mdframed}
\usepackage{hyperref}
\usepackage{cleveref}
\usepackage{booktabs}
\usepackage{url}

\usepackage{colortbl}
\usepackage{inconsolata}
\usetikzlibrary{arrows}

\usepackage{etex}
\newcommand{\todoc}[2]{{\textcolor{#1}{\textbf{#2}}}}

\definecolor{lightred}{RGB}{255, 204, 204}

\renewcommand{\todoc}[2]{\relax}

\newcommand{\newsubsubsection}[1]{\smallskip \noindent \textbf{\emph{#1}}}
\newcommand{\code}[1]{\texttt{\small #1}} %
\newcolumntype{a}{>{\columncolor{lightgray}}c} %

\newcommand{\ours}{\textsc{Waffle}\xspace}
\newcommand{\noattn}{\textsc{Waffle}$_\texttt{-attn}$\xspace}
\newcommand{\nocontra}{\textsc{Waffle}$_\texttt{-contra}$\xspace}

\newcommand{\attn}{structure-aware attention}

\newcommand{\testone}{WebSight-Test\xspace}
\newcommand{\testtwo}{Design2Code\xspace}
\definecolor{lightgray}{gray}{0.9}

\newcommand{\summary}[1]{
    \setlength{\parskip}{0pt} 
    \begin{mdframed}[linecolor=gray,roundcorner=12pt,backgroundcolor=gray!15,linewidth=3pt,innerleftmargin=2pt, leftmargin=0cm,rightmargin=0cm,topline=false,bottomline=false,rightline=false]
    \textbf{Summary:} #1
    \end{mdframed}
    \setlength{\parskip}{0pt}
}

\newcommand{\distance}{10pt}
\begin{document}

\title{\ours{}: Fine-tuning Multi-Modal Models for Automated Front-End Development
\thanks{In the Proceedings of ACL 2025. This version is for personal use only.}}

\author{{\normalsize Shanchao Liang} \\
{\normalsize Purdue University} \\
{\normalsize liang422@purdue.edu}
\And
{\normalsize Nan Jiang} \\
{\normalsize Purdue University} \\
{\normalsize jiang719@purdue.edu}
\And
{\normalsize Shangshu Qian} \\
{\normalsize Purdue University} \\
{\normalsize qian151@purdue.edu}
\And
{\normalsize Lin Tan} \\
{\normalsize Purdue University} \\
{\normalsize lintan@purdue.edu}
}

\maketitle

\begin{abstract}
Web development involves turning UI designs into functional webpages, which can be difficult for both beginners and experienced developers due to the complexity of HTML's hierarchical structures and styles. While Large Language Models (LLMs) have shown promise in generating source code, two major challenges persist in UI-to-HTML code generation: (1) effectively representing HTML's hierarchical structure for LLMs, and (2) bridging the gap between the visual nature of UI designs and the text-based format of HTML code.
To tackle these challenges, we introduce \ours, a new fine-tuning strategy that uses a structure-aware attention mechanism to improve LLMs' understanding of HTML's structure and a contrastive fine-tuning approach to align LLMs' understanding of UI images and HTML code. Models fine-tuned with \ours show up to 9.00 pp (absolute percentage point) higher HTML match, 0.0982 higher CW-SSIM, 32.99 higher CLIP, and 27.12 pp higher LLEM on our new benchmark \testone and an existing benchmark \testtwo, outperforming current fine-tuning methods.
\end{abstract}

\section{Introduction}
While Large Language Models have significantly advanced the automation of code generation in popular programming languages such as Python or Java~\cite{mistral,related_background2_llama,incoder,codegen,codellama,related8deepseekcoder,starcoder,starcoder2}, the automation of HTML code generation from UI design remains under-explored and challenging. As the core of front-end development, this task requires the model to understand not only the transformation from natural languages (NL) to programming languages (PL) but also from visual designs to PL. Recently, Multi-modal Large Language Models (MLLMs) have brought much progress in generating text from image descriptions~\cite{related6clip,related7siglip,lavis,llava,llava1.5,blip,blip2,instructblip,nougat,sharegpt4v,vary,moondream}. On top of this, a few MLLMs have been fine-tuned using UI image-to-code datasets (e.g., WebSight~\cite{related1websight}, Design2Code~\cite{related3design2code}).
Nonetheless, these approaches mainly apply standard fine-tuning and fail to address specific HTML code generation challenges.

\begin{figure}[t]
    \centering
    \includegraphics[width=\linewidth]{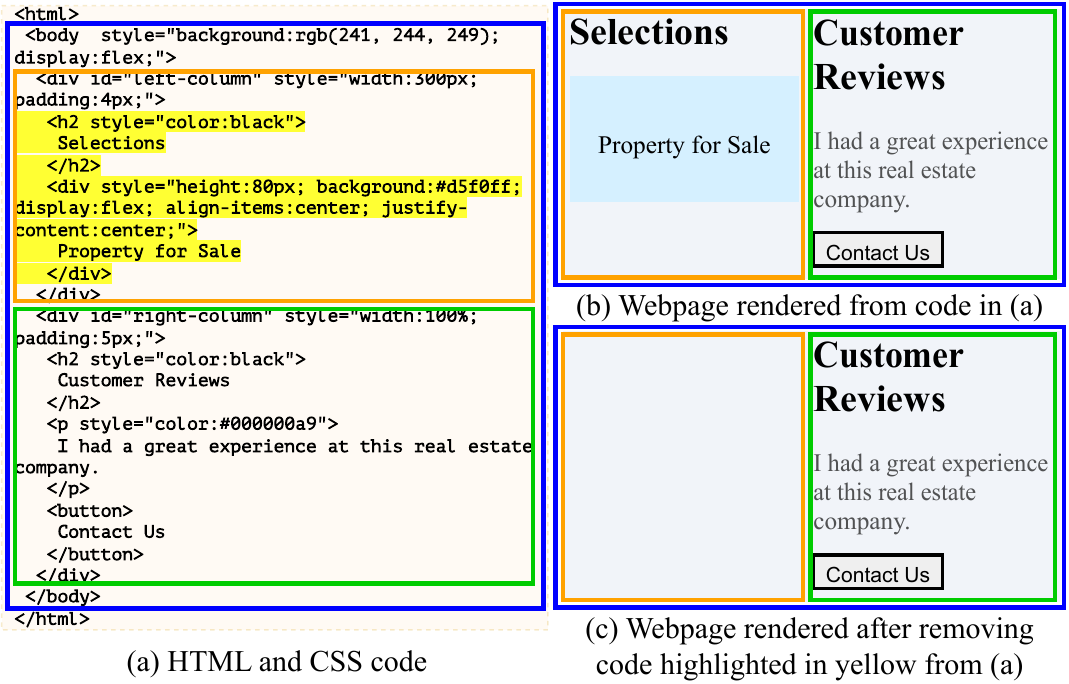}
    \caption{Removing the children of the element \code{<div id = "left-column">} highlighted in yellow does not affect the structure of the visual layout of itself or its sibling element \code{<div id="right-column">}. }
    \label{fig:attention-motivation}
\end{figure}

Two key challenges exist in translating UI design images to HTML code: (1) how to teach models to learn effectively the \textbf{domain knowledge of  HTML structures}, which significantly influences the rendered UI design, and (2) how to teach the models to learn the \textbf{subtle differences in the visual understanding of UI images and text understanding of HTML code}. 

Regarding the first challenge, there are three basic structural aspects of HTML code. Firstly, all the styles of the parent element are directly passed to the children unless specifically overridden. Secondly, the layout of the siblings affects each other. Thirdly, nodes are not affected by the subtrees of their siblings. The last principle might be less obvious compared to the first two, and we explain it with an example. \Cref{fig:attention-motivation} shows (a)  an HTML code file and (b)  its rendered webpage. We use blue, orange, and green blocks to map the code chunks and their corresponding visual rendering on the webpage. The top-level \code{<body>} element refers to the whole webpage, the child element \code{div id="left-column"} refers to the left part of the webpage, and another child element \code{div id="right-column"} refers to the right part. Modifications to the child of \code{div id="left-column"} do not change how the \code{div id="right-column"} looks on the webpage (even if we remove all the content inside \code{div id="left-column"} as (c) shows).

To learn such domain knowledge of HTML code structure, we propose a novel \textbf{structure-aware attention} mechanism. The structure-aware attention captures the structure information of the HTML code by explicitly allowing tokens to focus on three types of previous code segments that are the most relevant (details in ~\Cref{sec:attention}). With such structural information in the HTML code, \ours can focus on parts of the code that have the most influence on the resulting UI design, thus benefiting the overall performance.

\begin{figure*}[t]
    \centering
    \includegraphics[width=0.95\linewidth]{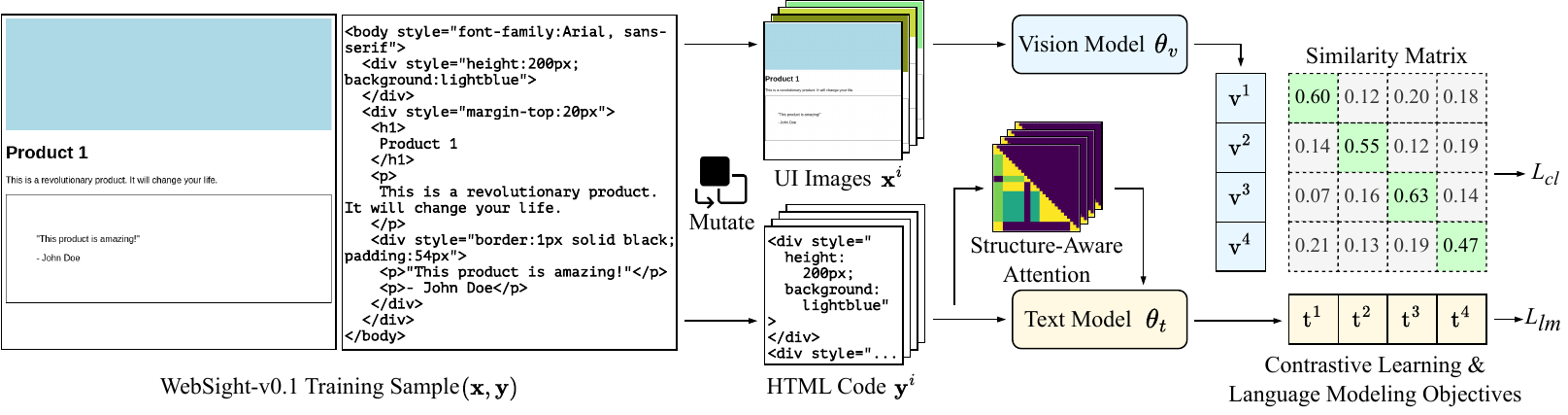}
    \caption{Overview of \ours, including training data mutation, structure-aware attention, and contrastive learning.}
    \label{fig:overview}
\end{figure*}

For the second challenge, minor variations in layout structure can still impact the content of the generated code. However, MLLMs often fail to capture these subtle differences and generate identical code for visually distinct inputs. We illustrate this issue in ~\Cref{challeng2_example}. To enable MLLMs to recognize subtle differences in UI images due to minor changes in the code, we adopt \textbf{contrastive learning}~\cite{related7siglip, related6clip, related9constrastive-simcse} to the current task to teach MLLMs to focus on important visual differences. 
 
Combining the two approaches, this paper introduces \ours, a fine-tuning pipeline specifically designed for UI images to HTML code generation, with the following contributions:
\begin{enumerate}[leftmargin=15pt, noitemsep, topsep=0pt]
    \item We design \emph{structure-aware attention} for HTML code generation, which enables MLLMs to learn the structure knowledge of HTML code.
    \item We apply \emph{contrastive learning} algorithms to boost MLLMs' understanding of the details in the rendered UI images and the HTML code. 
    \item We create a new dataset of 231,940 pairs of webpages and their HTML code, which could facilitate future research on web MLLMs. 
    \item We conduct comprehensive experiments on two backbone MLLMs. \ours{} improves the backbone MLLMs by achieving up to 9.00 pp higher HTML Match, 0.0982 higher CW-SSIM, 32.99 higher CLIP, and 27.12 pp higher LLEM.
    \item We highlight that \ours{} as a fine-tuning approach, is model-independent and can be applied to improve any MLLMs for UI-to-HTML code generation.
\end{enumerate}
\noindent \textbf{Availability:} \url{https://github.com/lt-asset/Waffle}

\section{Approach}
\noindent\Cref{fig:overview} represents the overview of  \ours.  We create a new mutated HTML dataset (\Cref{subsec:contrastive_dataset}) for training and fine-tuning. In addition, we design structure-aware attention (\Cref{sec:attention}) during model fine-tuning to teach models to focus on important HTML segments. Finally, we use contrastive learning training to teach models to learn  visual differences  and align the models' visual and HTML/CSS code understanding 
(\Cref{sec:contrastive_learning}). 
\emph{We note that \ours is a generalizable fine-tuning pipeline that can benefit any pre-trained MLLMs.}

Specifically, we construct the training dataset by applying mutation rules for HTML code on a subset of a popular dataset, WebSight-v0.1, to generate the corresponding source code and UI images.

\begin{table}[h]
\centering
\scriptsize
\setlength{\tabcolsep}{4.7pt}
    \begin{tabular}{cccccc|c|c}
    \toprule
    \multicolumn{6}{c|}{\textbf{CSS}} & \multirow{2}{*}{\textbf{HTML}} & \multirow{2}{*}{\textbf{Total}}\\
    Color & Size & Margin & Font  & Display & Position &  & \\ 
    \midrule
    12 & 11 & 19 & 10 & 1 & 2 & 8 & 63 \\
    \bottomrule
    \end{tabular}
\caption{Most frequent causes of failures.}
\label{table:fault_analysis}
\end{table}

\subsection{Training Data Mutation}
\label{subsec:contrastive_dataset}

\noindent To teach MLLMs the important visual differences, 
we create \ours{}'s contrastive training data from WebSight-v0.1, which is a fine-tuning dataset built by HuggingfaceM4 and contains 822,987 pairs of HTML code and its corresponding screenshots~\cite{related1websight}. However, to make contrastive learning effective, the model must learn to recognize subtle yet realistic variations in UI elements.

To achieve this, we analyze common mistakes in VLM-WebSight by conducting a failure analysis on 50 validation samples, identifying seven common error categories (\Cref{table:fault_analysis}). Using insights from these failures, we design realistic mutation rules to modify HTML/CSS in WebSight-v0.1, ensuring that the contrastive training data reflects real-world errors. The mutation process follows the observed error frequencies; for example, since color mismatches account for 19.05\% of errors (12/63), mutations on color properties are applied at the same rate (Table 1). The full set of mutation rules is detailed in Appendix~\ref{sec: appendix_mutation_rule}.

Based on the mutation rules, we randomly sample 100,000 instances from the WebSight-v0.1 dataset, creating four distinct mutants with each sample. The mutation rules are applied based on the frequency of failures computed from the validation set. 
The final mutated dataset (after removing rendering failures, identical mutants, and blank images) has 57,985 groups, each group containing four pairs of HTML code and the corresponding rendered webpage images.

\subsection{Structure-Aware Attention}
\label{sec:attention}

\noindent HTML code has clear structures, and certain structural properties can be directly reflected in the rendered UI design.
Such domain knowledge can benefit the generation process of MLLMs. 
There are three most important elements for rendering an element's layout: its parent element, sibling elements, and the element itself. 
\ours{} implements a novel attention mask that provides each element with a pruned view of all previous tokens, including \textit{parent-attention}, \textit{sibling-attention}, and \textit{self-attention}. These attention masks allow the tokens to pay specialized attention to their parent elements, sibling elements, and themselves. 

\Cref{fig:hierarchical_attention} shows a simple example of \ours{}'s \attn{}. \Cref{fig:hierarchical_attention} (a) shows an HTML code snippet and (b) shows the DOM tree of HTML code in (a), where the root node is the \code{<body>} element. \code{<div id="leftCol">} and \code{<div id="rightCol">} are two children of node \code{<body>}, and they are siblings to each other. \code{<div id="leftCol">} has one child, the text \code{Selections}. \code{<div id="rightCol">} has one child, which is a \code{<h2>} element with the text \code{Customer Reviews} inside.

According to the domain knowledge that an element is mostly affected by its parent and sibling elements, \ours{} builds the \attn as shown in (c).

\newsubsubsection{Parent-Attention.} The parent-attention is from each element's tokens to its parent element's tokens.
\ours utilizes the fact that all the children elements inherit the parent element's styles and structure.
For instance, the tokens of the element \code{<div id="leftCol">} pay parent-attention to tokens of the element \code{<body>}, and the tokens of the element \code{Selections} pay parent-attention to tokens of the element \code{<div id="leftCol">}. 

\newsubsubsection{Sibling-Attention.} The sibling attention is from each element's tokens to its preceding sibling elements' tokens. \ours{} utilizes the fact that sibling elements under the same parent can affect the arrangement and style of each other, so each element needs to pay attention to its preceding siblings. For instance, the tokens of the element \code{<div id="rightCol">} pay sibling-attention to tokens of the element \code{<div id="leftCol">}.

\newsubsubsection{Self-Attention.} This is the standard self-attention mechanism, it allows each token to focus on all previous tokens within the same element, excluding the children elements. To illustrate, all tokens in a specific element have self-attention (yellow cells~\Cref{fig:hierarchical_attention}) to all the tokens belonging to the element itself.

\ours{} applies the \attn{} mechanism exclusively to the language model decoder while keeping the vision encoder unchanged. Specifically, \ours{} applies the structure-aware attention mask—formed by the union of parent-attention, sibling-attention, and self-attention masks—to only one-fourth of the attention heads in the decoder. This enables these heads to explicitly capture structural domain knowledge, while the remaining three-fourths retain full self-attention and leverage pre-trained knowledge. The number of attention heads using \attn{} is a tunable hyperparameter, which can be adjusted as described in~\Cref{sec:ablation_study}.

\definecolor{mypurple}{RGB}{128, 0, 128} %

\begin{figure*}[htp]
    \centering
    \includegraphics[width=0.9\linewidth]{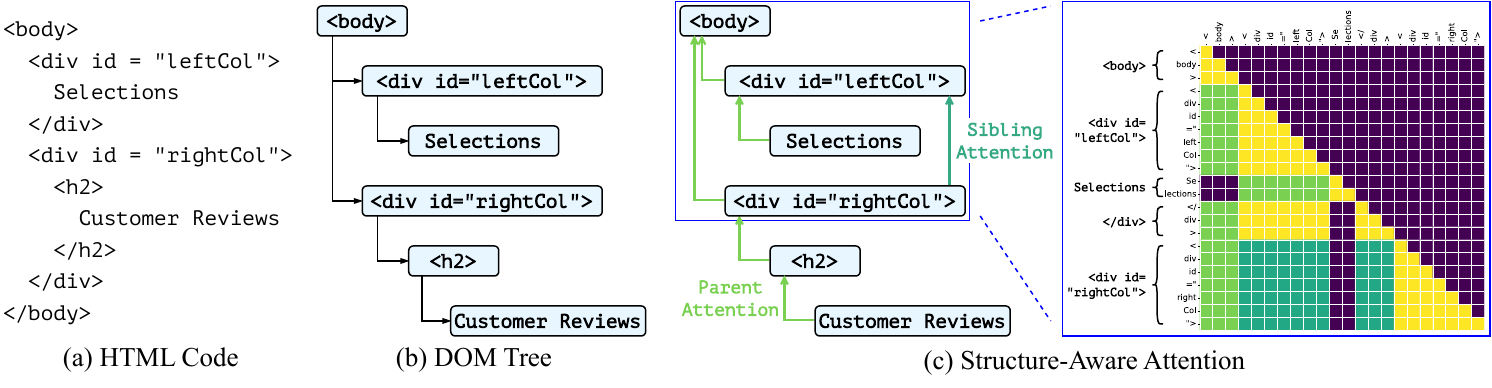}
    \caption{Example of structure-aware attention. \colorbox{yellow}{\raisebox{0pt}[5pt][0pt]{Yellow}}: self-attention.  \colorbox{green}{\raisebox{0pt}[5pt][0pt]{Green}}: parent-attention.  \colorbox{teal!60}{\raisebox{0pt}[5pt][0pt]{Teal}}: sibling-attention.  \colorbox{mypurple!60}{\raisebox{0pt}[5pt][0pt]{Purple}}: masked-out Attention.}
    \label{fig:hierarchical_attention}
\end{figure*}

\subsection{Contrastive-Learning}
\label{sec:contrastive_learning}

To address current models' limited ability to effectively map variations in HTML code to the corresponding UI, \ours utilizes contrastive learning, allowing models to learn from comparisons and contrasts between similar examples.

More concretely, the training dataset consists of groups of HTML codes and UI Image pairs, where each group has $k$ such pairs. During training, in any group, for image $\mathbf{x}^i$ and code $\mathbf{y}^i$ ($i \in \{1, ..., k\}$, image $\mathbf{x}^i$ is the rendered webpage image from code $\mathbf{y}^i$),
the webpage image is split into a list of pixel patches. Each patch is encoded by the vision model $\theta_v$ and fused to the text model's latent space using an adapter to result in a list of patch embeddings $\mathbf{v}^i = \{v^i_1, v^i_2, ..., v^i_{M}\}$ ($M$ is a hyper-parameter of the backbone MLLM).
The HTML code is tokenized into tokens $\mathbf{y}^i = \{y^i_1, y^i_2, ..., y^i_{N^i}\}$, where $N^i$ is the number of the tokens of the HTML code $\mathbf{y}^i$. These tokens are encoded to embeddings $\mathbf{t}^i = \{t^i_1, t^i_2, ..., t^i_{N^i}\}$ by the language model $\theta_t$ (using the \attn{}).
We use the average over all the patch embeddings to represent the embeddings of the webpage image, and the average over all the tokens' embeddings to represent the embedding of the whole HTML code, denoted by:

{
\setlength\abovedisplayskip{0pt}
\setlength\belowdisplayskip{0pt}
\footnotesize
\begin{equation}
    \overline{v^i} = \frac{1}{M}\sum_{j=1}^{M} v^i_j \;\;, \quad \overline{t^i} = \frac{1}{N^i}\sum_{j=1}^{N^i} t^i_j
\end{equation}
}

Then we calculate the cosine similarity score between the code and image embeddings, $\overline{t^i}$ and $\overline{v^i}$, bipartitely. 
The contrastive learning objective is to maximize the similarity between the embeddings of the corresponding code and UI, by minimizing the contrastive learning loss $L_{cl}$, along with the language modeling loss $L_{lm}$ as follows:

{
\setlength\abovedisplayskip{0pt}
\setlength\belowdisplayskip{0pt}
\footnotesize
\begin{equation}
    L_{cl} = -\sum_{i=1}^k \log \left( \frac{\mathop{exp}\left (\mathop{sim}{(\overline{t^i}, \overline{v^i})}\right )}{\sum_{j=1}^k \mathop{exp}\left (\mathop{sim}{(\overline{t^i}, \overline{v^j})}\right )}\right)
\end{equation}
\begin{equation}
    L_{lm} = -\sum_{i=1}^k\sum_{j=1}^{N_i} \log P(y^i_j \mid \mathbf{y}^i_{<j}, \mathbf{x}^i, \theta_t, \theta_v)
\end{equation}
}

The contrastive learning loss aims to maximize the similarity scores at the diagonal of the similarity matrix (as the green cells of the similarity matrix shown in ~\Cref{fig:overview}). 
{It trains the MLLM's vision model, $\theta_v$, which encodes a webpage, $\mathbf{x}^i$, to closely match the encoded embeddings of the text model, $\theta_t$, on the corresponding HTML code, $\mathbf{y}^i$.}

The language modeling loss, on the other hand, aims to maximize the probability of generating the correct token $y^i_j$ given all previous token $\mathbf{y}^i_{<j}$ and the input webpage image $\mathbf{x}^i$. This is the standard objective of training an MLLM to generate text from images.
\ours jointly optimizes the two objectives, {\small $L_{\ours} = L_{lm} + \lambda L_{cl}$},
\noindent where $\lambda$ is a hyper-parameter constant controlling the effect of contrastive learning on the overall optimization.

\section{Experimental Setup}

\subsection{Model Training}
\label{sec:model_training}
\noindent
We implement \ours on two backbones, VLM-WebSight, and Moondream2~\cite{related1websight,moondream}. Each backbone is first fine-tuned on the WebSight-v0.1 dataset using the standard language modeling objective. For VLM-Websight, we use the released fine-tuned checkpoint with details in~\cite{related1websight}. This checkpoint is fine-tuned using DoRA~\cite{dora} (a variant of parameter-efficient training, LoRA~\cite{lora}, with \code{rank} set to 64). 
For Moondream2, we also fine-tune the model using DoRA~\cite{dora} (with \code{rank} set to 64, and \code{lora\_alpha} set to 128).
The model's weights are updated using the AdamW~\cite{adamw} optimizer, with the learning rate set to $3e^{-5}$. The batch size is 64.

On top of the fine-tuned MLLMs from the first step, we apply the structure-aware attention and contrastive learning approach. Structure-aware attention is applied to $\frac{1}{4}$ of the attention heads in each attention layer in the LLM decoder. Each model is trained with DoRA using the combined learning objective on the contrastive learning dataset ($\lambda$ set to 0.1). The model's weights are updated using the AdamW optimizer, with the learning rate set to $2e^{-5}$. The batch size is 32.

\subsection{Test Data} 
We evaluate \ours{} using two test datasets: \testone{}, which consists of synthetic webpages, and \testtwo{}, which consists of real-world webpages. Since WebSight-v0.1 was already used for training, we created \testone{} following the same
process as WebSight-v0.1~\cite{related1websight}. In total, \testone{} contains 500 test samples, each has a webpage image and the respective ground-truth HTML source code.

\testtwo{} is an open-source benchmark of 484 manually processed real-world website screenshots, which are more complex than Websight-v0.1. Evaluation on \testtwo{} indicates
the generalization ability of  
models fine-tuned on \ours's training dataset 
to real-world scenarios.

\subsection{Evaluation Metrics}
\label{sec:metrics}

We note that existing methods struggle to effectively capture subtle or structural differences in images. To address this, we adopt CLIP and Low-Level Element Matching (LLEM) metrics from previous work~\cite{related3design2code}. Additionally, we introduce two new metrics: HTML-Match, designed to evaluate structural similarity in rendered HTML, and complex wavelet structural similarity index (CW-SSIM) to detect minor differences in images, respectively.

\newsubsubsection{HTML-Match.}
HTML-Match is the percentage of generated images that match the ground truth images perfectly at the pixel level. For this comparison, styles and attributes are removed from both the ground truth and generated HTML.
This process emphasizes the model's ability to accurately recognize the text content and the DOM tree structure of the HTML code, and assess MLLM's ability to recover the HTML structure from the given UI.

\begin{figure}[t]
    \centering
    \includegraphics[width=1\linewidth]{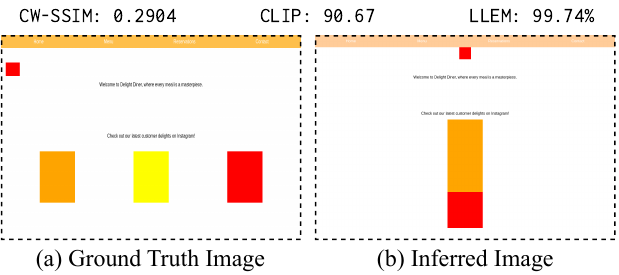}
    \caption{Different metrics on a pair of images.}
    \label{fig:cwssim_example}
\end{figure}

\newsubsubsection{CW-SSIM.}
CW-SSIM computes the structural similarity between images~\cite{metrics-cwssim}. In certain cases, LLEM and CLIP scores might yield high scores as they failed to capture the structural difference. For example, in \Cref{fig:cwssim_example} the rendered image is structurally different from the ground truth image, resulting in low CW-SSIM scores, 0.2904. However, the LLEM gives a high score of 99.74\%, and the CLIP score is 90.67, which does not align with human evaluation as well as CW-SSIM. 

\newsubsubsection{CLIP.}
CLIP score~\cite{related6clip, related3design2code}  measures the similarity between the rendered webpage of the inferred HTML code and the ground truth webpage based on the \code{CLIP-ViT-B/32}'s embeddings of webpages. 

\newsubsubsection{Low-Level Element Matching (LLEM).} Previous work~\cite{related3design2code} proposes LLEM to measure the percentage of matched (1) text blocks, (2) text content, (3) position of each matched text block, and (4) font color within each text block.

In this work, we prioritize CW-SSIM as a more robust evaluation metric. Yet, we still report CLIP and LLEM as they are used in existing work.

\newsubsubsection{Human Evaluation.}
Two human annotators are asked to compare the rendered webpages generated by different techniques from a subset of test data with the ground-truth webpage and rank them.

\begin{table*}[htb!]
\centering
\scriptsize
\setlength{\tabcolsep}{2pt}
\begin{tabular}{llcccccccc}
\toprule
    \multirow[c]{3}{*}[0.5ex]{Backbones} &  \multirow[c]{3}{*}[0.5ex]{Techniques} & \multicolumn{4}{c}{\testone} && \multicolumn{3}{c}{\testtwo} \\
\cmidrule{3-6}\cmidrule{8-10}
     & & HTML-Match (\%) $\uparrow$ & CW-SSIM $\uparrow$ & CLIP $\uparrow$ & LLEM (\%) $\uparrow$ && CW-SSIM $\uparrow$ & CLIP $\uparrow$ & LLEM (\%) $\uparrow$ \\

\midrule \midrule
    Gemini 1.5 Pro & Prompting & 9.40  & 0.3385 & 88.55 & 90.16 && 0.2652  & 87.76 & 87.17 \\
    GPT-4o mini & Prompting & 10.20  & 0.3055 & 87.72 & 87.54 && 0.2304  & 86.06 & 78.84 \\
    GPT-4o& Prompting & 11.40 &  0.3666 & 89.03 &  92.18 && 0.2776 & 83.67 & 75.98\\
    
\midrule
\midrule
    \multirow{2}{*}{Moondream2} & Standard FT & 21.60  & 0.4233 & 89.92 & 90.59 &  & 0.1348 & 46.63 & 40.71 \\
    & \ours{} & \textbf{27.60} & \textbf{0.4486} & \textbf{89.98} & \textbf{91.72} &  & \textbf{0.2142}  & \textbf{79.62} & \textbf{67.83} \\
\midrule
    \multirow{2}{*}{VLM-WebSight} & Standard FT & 28.00  &  0.5023 & 93.30 & 92.73 && 0.2518 & 82.35 & 73.00 \\
    & \ours & \textbf{37.00}   &  \textbf{0.6005}  & \textbf{94.57} & \textbf{95.16} && \textbf{0.2815} & \textbf{85.98}  & \textbf{77.81}   \\
\bottomrule
\end{tabular}\\
\vspace{2pt}
\hspace{8em}\scriptsize{*Gemini 1.5 Pro's results are on 384 test instances as it generates no answers for the remaining 100 instances.}
\caption{Main results on the \testone and \testtwo dataset.}
\label{table:main_waffle_design}
\end{table*}

\begin{figure*}[t]
    \centering
    \includegraphics[width=\linewidth]{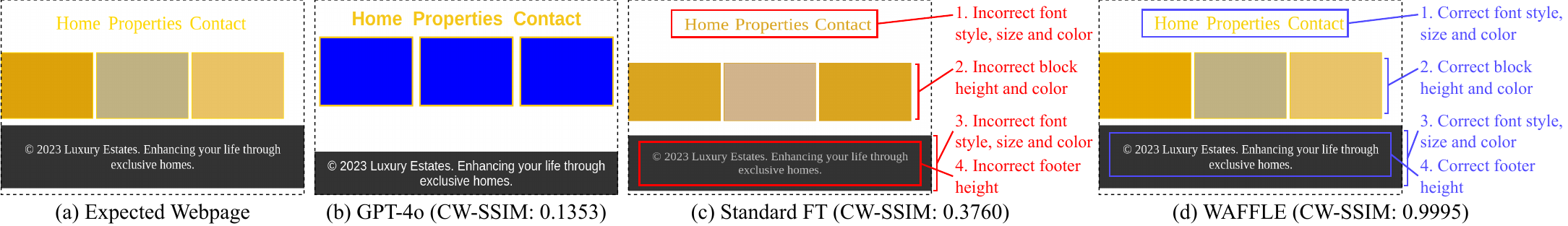}
    \caption{Example from \testone dataset, with the generated images by GPT-4o, Standard FT (VLM-WebSight), and \ours (VLM-WebSight).}
    \label{fig:rq1_case_study}
\end{figure*}

\section{Results}
\subsection{Effectiveness of \ours}
\label{sec:rq1}
 
\noindent \Cref{table:main_waffle_design} show the performance of various fine-tuning strategies on the two testing datasets, \testone (500 samples) and \testtwo (484 samples). 

\newsubsubsection{Comparison against standard fine-tuning.}
We compare \ours with the baseline, the standard fine-tuning (FT) method. In both tables, \ours achieves significant improvements over the standard FT on 
\emph{\textbf{all metrics}} with Moondream2 and VLM-WebSight as the backbone. 

On \testone, \ours achieves 6.00 pp higher HTML-Match (27.60\% vs. 21.60\%) and 0.0253 higher CW-SSIM (0.4486 vs. 0.4233) 
than Standard FT with Moondream2 as the backbone. On the VLM-WebSight backbone, \ours{} outperforms Standard FT, i.e., original VLM-WebSight, by 9.00 pp HTML-Match, 0.0982 CW-SSIM, improving the model's ability to generate structurally similar images and align the two modalities.
On \testtwo, \ours achieves greater improvements compared to the Standard FT technique with both backbones.

\summary{
Overall, \ours significantly improves all metrics for both backbones in both test datasets over standard fine-tuning with up to 9.00 pp for HTML Match, 0.0982 for CW-SSIM, 32.99 for CLIP, and 27.12 pp for LLEM.
}

\newsubsubsection{Comparison against SOTA commercial models.} 
Due to the lack of comparable baselines, we compare \ours with top commercial models, which include GPT-4o mini, GPT-4o, and Gemini 1.5 Pro. We apply direct prompting following prior work~\cite{related3design2code}. The result is shown in~\Cref{table:main_waffle_design}.

On the \testone dataset, models fine-tuned with \ours perform better than the SOTA commercial models. VLM-WebSight outperforms GPT-4o by 25.60 pp on HTML-Match (37.00\% vs. 11.40\%) and 0.2339 on CW-SSIM (0.6005 vs. 0.3666), showing \ours{'s} benefit in addressing the two challenges, i.e., generate structurally correct HTML and closing the gap between the two modalities. Similarly, for the smaller backbone, Moondream2 exceeds GPT-4o on CW-SSIM and HTML-Match.

As shown in~\Cref{table:main_waffle_design}, VLM-WebSight fine-tuned by \ours is better than GPT-4o on CW-SSIM by 0.0039, and better than GPT-4o mini by 0.0511 on the \testtwo dataset. However, GPT-4o is better in the other two metrics versus VLM-WebSight.
Moondream2 fine-tuned by \ours has a lower performance compared to GPT-4o and GPT-4o mini on all metrics. This is likely due to its smaller size, which could influence its generalizability to more complex, out-of-distribution data. %

\summary{
On simpler data, \ours{} achieves better or comparable results than SOTA commercial models, with 16.20--25.60 pp improvement on HTML-Match and 0.0820--0.2339 improvement on CW-SSIM. On more complex data, \ours enables VLM-WebSight to outperform commercial models on CW-SSIM.
}

\begin{table*}[htb!]
\centering
\scriptsize
\setlength{\tabcolsep}{2pt}
\begin{tabular}{llcccccccc}
\toprule
    \multirow[c]{3}{*}[0.5ex]{Backbones} &  \multirow[c]{3}{*}[0.5ex]{Techniques} & \multicolumn{4}{c}{\testone} && \multicolumn{3}{c}{\testtwo} \\
\cmidrule{3-6}\cmidrule{8-10}
     & & HTML-Match (\%) $\uparrow$ & CW-SSIM $\uparrow$ & CLIP $\uparrow$ & LLEM (\%) $\uparrow$ && CW-SSIM $\uparrow$ & CLIP $\uparrow$ & LLEM (\%) $\uparrow$ \\
\midrule
\midrule
    \multirow{4}{*}{Moondream2} & Standard FT & 21.60  & 0.4233 & 89.92 & 90.59 && 0.1348 & 46.63 & 40.71 \\
    & \noattn & 23.60  &  0.4311 & \textbf{90.47} & 91.34 && 0.1821  & 67.73 & 56.49 \\
    & \nocontra & 26.00 & 0.4296 & 89.55 & 91.21 && 0.2100  & 76.63 & 65.82 \\
    & \ours{} & \textbf{27.60} & \textbf{0.4486} & 89.98 & \textbf{91.72} && \textbf{0.2142}  & \textbf{79.62} & \textbf{67.83} \\
\midrule
    \multirow{4}{*}{VLM-WebSight} & Standard FT & 28.00  &  0.5023 & 93.30 & 92.73 && 0.2518 & 82.35 & 73.00 \\
    & \noattn & 30.80  &  0.5411 & 94.29 & 94.20 && 0.2480 & 85.64 & 75.34 \\
    & \nocontra & 35.80  & 0.5677 & \textbf{95.08} & \textbf{95.30} && 0.2653 & 85.16 & 76.48 \\
    & \ours & \textbf{37.00} & \textbf{0.6005}  & 94.57 & 95.16 && \textbf{0.2815} & \textbf{85.98}  & \textbf{77.81} \\
\bottomrule
\end{tabular}
\caption{Ablation studies on the two test datasets. LLEM refers to the averaged Low-Level Element Matching.}
\label{table:rq2}
\end{table*}

\subsection{Case Study}
\label{sec: appendix_case_study}

Figure~\ref{fig:rq1_case_study} shows the generation results for one instance from \testone. The generated webpage of GPT-4o has a CW-SSIM of 0.1353, significantly lower than that of 0.3760 from VLM-WebSight under standard fine-tuning. On the other hand, the webpage generated by VLM-WebSight fine-tuned by \ours reaches an almost perfect CW-SSIM score. This example shows the effectiveness of using \ours on UI-to-HTML generation.

\subsection{Ablation Studies}
\label{sec:ablation_study}

\newsubsubsection{Comparison against ablation models.} 
We compare \ours with two ablation models:

\begin{itemize}[leftmargin=2ex, topsep=0pt, noitemsep]
    \item \noattn: This is \ours with only contrastive learning and without the use of structure-aware attention. 
    \item \nocontra: This is \ours with only structure-aware attention. 
\end{itemize}

Shown in \Cref{table:rq2},
\noattn brings improvements compared to the Standard FT on all metrics, and \nocontra brings a 4.40 pp improvement on HTML-Match. 
On the \testtwo dataset, \ours has dominating performance on \textbf{\emph{all metrics}} for models fine-tuned with both backbones. 
Across the two backbones, models fine-tuned with \ours are higher than those fine-tuned with \noattn by up to 0.0335 on CW-SSIM.

\summary{
Contrastive learning and structure-aware attention significantly improve performance over standard fine-tuning. On the simpler \testone data, models trained with \ours achieve the highest HTML-Match and CW-SSIM scores. On the more complex \testtwo data, \ours still delivers the best results across all metrics for both backbones.
}

\begin{table}[t]
\centering
\scriptsize
\setlength{\tabcolsep}{3pt}

\begin{tabular}{lcaca}
\toprule
   {Techniques} & Rank 1 $\uparrow$ & Rank 2 $\uparrow$ & Rank 3 $\uparrow$ & Avg Rankings $\downarrow$ \\
\midrule
Standard FT & \phantom{0}9 \textbar\ 34 \scriptsize{(43)} & 18 \textbar\ 13 \scriptsize{(46)} & 34 \textbar\ 25 \scriptsize{(54)} & 2.88 \textbar\ 2.37 \scriptsize{(2.62)} \\

\noattn & 22 \textbar\ 23 \scriptsize{(45)} & \phantom{0}11 \textbar\ 30 \scriptsize{(41)} & 22 \textbar\ 26 \scriptsize{(71)} & 2.67 \textbar\ 2.45 \scriptsize{(2.56)} \\

\nocontra & 60 \textbar\ 33 \scriptsize{(93)} &  \phantom{0}13 \textbar\ 17 \scriptsize{(30)} & 11 \textbar\ 26 \scriptsize{(42)} & 1.78 \textbar\ 2.41 \scriptsize{(2.10)} \\

\ours & 46 \textbar\ 54 \scriptsize{(\textbf{100})} &  33 \textbar\ 22 \scriptsize{(55)} & 10 \textbar\ \phantom{0}15 \scriptsize{(26)} & 1.85 \textbar\ 1.79 \scriptsize{(\textbf{1.82})} \\
\bottomrule
\end{tabular}
\caption{Human evaluation on two datasets using VLM-WebSight as the backbone. The numbers are shown as ``\texttt{x}\textbar \texttt{y} (x+y)'', where \texttt{x} is the result on \testone and \texttt{y} is the result on \testtwo.}
\label{table:ablation_gold_medals_one}
\end{table}

\newsubsubsection{Human evaluation results.}
We select 50 test samples from \testone and \testtwo (100 total). Each sample has four generated HTML codes and rendered webpages from Standard FT, \noattn, \nocontra, and \ours. Human raters rank the generated webpages based on similarity to the ground-truth webpage without knowing which model produced each one. Multiple results can share the same rank if they are deemed equally similar to the ground truth. \Cref{table:ablation_gold_medals_one} shows the human evaluation results for VLM-WebSight fine-tuned by Standard FT, \noattn, \nocontra, and \ours.

Across both datasets, \ours has the best averaged rankings, 1.82, outperforming both ablations and the baseline. Specifically, \ours reaches 54 times rank 1 placement on \testtwo, showing great generalizability on more complex datasets. \nocontra is the second best technique on the two testsets, reaching 93 times rank 1, and an average ranking of 2.10. On the other hand, \noattn is the third-best technique but still outperforms standard FT.

\summary{
Human evaluation shows that (1) both structure-aware attention and contrastive learning contribute to the code generation quality, and (2) \ours-generated HTML/CSS code is consistently rated higher than code generated with standard fine-tuning. 
}

\subsection{Structure-Aware Attention's Effect}
\label{subsec:attn_effect}

\begin{table}[t]
\centering
\scriptsize
\setlength{\tabcolsep}{12pt}
\begin{tabular}{lrrr}
\toprule
   {Techniques} & Prior & Current & Drop (\%) \\
\midrule
\noattn &  0.8002 & 0.5797 & 27.55 \\
\ours &  0.8291 & 0.7932 & 4.34 \\
\bottomrule
\end{tabular}
\caption{
CW-SSIM on 20 samples using the VLM-WebSight backbone. ``Prior'' refers to ``without intermediate mistakes'', and ``Current'' to ``with intermediate mistakes''.
}
\label{table:rq3_result}
\end{table}

To demonstrate how \attn{} helps MLLMs focus on the correct structural elements (such as parent and sibling elements) during generation, we introduce controlled errors in the generation process. Specifically, we select 20 high-performing samples—those with the highest CW-SSIM scores—generated by VLM-WebSight fine-tuned with \ours{} and \noattn{}. These samples indicate cases where the models originally performed well on UI-to-code generation. Our goal is to evaluate whether models fine-tuned with \attn{} exhibit greater robustness in handling intermediate errors compared to those without it.

For each selected sample, we manually modify the partially generated HTML code by editing the sibling elements of the correct structure, following the example in~\Cref{fig:attention-motivation}. These modifications alter the rendered output, simulating realistic errors that could occur during generation. The models are then tasked with re-completing the HTML code while being provided the same UI image with the partially generated, incorrect HTML as input. Since the modifications affect sibling elements rather than the primary structure, an ideal model should recover from these errors without cascading failures in subsequent generations.

\Cref{table:rq3_result} shows the results of re-completion following the intermediate mistakes. With \noattn, the averaged CW-SSIM across the 20 samples drops by 0.2205 (from 0.8002 to 0.5797) if the model makes intermediate mistakes. By contrast, 
with \ours, the averaged CW-SSIM only drops by 0.0359, from 0.8291 to 0.7932.

\summary{
Integrating structure-aware attention brings stability to model generation, making models more robust against 
intermediate mistakes, and reducing the performance drop by 23.31\%, from 27.55\% to 4.34\%, ensuring more consistent generation quality. }

\subsection{Contrastive Learning's Effect}

\noindent{}To show contrastive learning's effect on MLLMs' visual and textual understanding, we design two experiments. The first experiment analyzes whether MLLM's understanding of the image and code is aligned through the integration of contrastive learning, and the second experiment analyzes whether contrastive learning can teach the model to capture the subtle difference in the images. 

\newsubsubsection{Aligning models' two modalities.}
Specifically, under the same procedure, we compute the averaged text embeddings and image embeddings for a subset (12 samples from \testone dataset) of the test samples in~\Cref{subsec:attn_effect} using the Moondream2 model fine-tuned by \noattn and Standard FT. Then for each pair of the averaged image embeddings and text embeddings, $(\overline{v^i}, \overline{t^i})$, we normalize them and compute the Euclidean distance and the cosine similarity between them. 

\Cref{table:rq4_text2image} shows the results of the measurements for both techniques. The Euclidean distance between the embeddings of the two modalities is 0.8447 for \noattn, which is lower than 1.3395, the distance of the embeddings from Standard FT, by 0.4948. Similarly, the cosine similarity between the embeddings encoded by \noattn is higher than the Standard FT by 0.5217 (0.6244 vs. 0.1027).

\begin{table}[t]
\centering
\scriptsize
\setlength{\tabcolsep}{8pt}
\begin{tabular}{lccc}
\toprule
   {Techniques} & $d(\overline{v^i}, \overline{t^i}) \downarrow$ & $sim(\overline{v^i}, \overline{t^i}) \uparrow$ \\
\midrule
Standard FT &  1.3395 & 0.1027 \\
\noattn &  0.8447 & 0.6244 \\
\bottomrule
\end{tabular}
\caption{Distance ($d$) and similarity ($sim$) between averaged image embeddings $\overline{v^i}$ and text embeddings $\overline{t^i}$, using Moondream2 as the backbone.}
\label{table:rq4_text2image}
\end{table}

\begin{figure}[t]
    \centering
    \includegraphics[width=\linewidth]{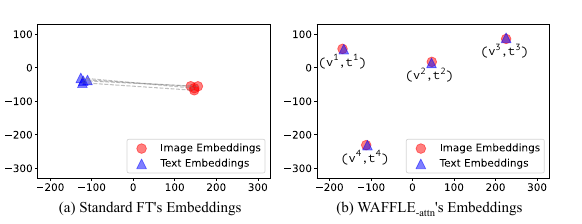}
    \caption{t-SNE plots of the text and image embeddings, computed by Moondream2 fine-tuned with Standard FT and \noattn.}
    \label{fig:case_study_text_and_image}
\end{figure}

In addition, \Cref{fig:case_study_text_and_image} demonstrates that contrastive learning teaches the model to align the text and image understandings. The vision embeddings (red circles) are far away from their corresponding text embeddings (blue triangles) when calculated by Standard FT. By contrast, the vision embeddings are grouped with their corresponding text embeddings by \noattn.

\summary{Fine-tuning with contrastive learning (\noattn) enhances MLLM visual-textual alignment, making image and code embeddings more cohesive, reducing Euclidean distance while elevating the cosine similarity between them. }

\section{Related Work}

\subsection{Multi-Modal Large Language Models}
Recent advances in vision and language models have greatly improved MLLMs' capabilities in tasks like image captioning~\cite{related7siglip, blip,blip2, idefmics}, text-to-image generation~\cite{related-dalle2,related-stablediffusion}, visual question answering~\cite{llava,llava1.5, Qwen-VL}, and document editing~\cite{related-docedit}. While popular models like Llava~\cite{llava,llava1.5}, Qwen-VL~\cite{Qwen-VL}, and Vary~\cite{vary} perform well in general image tasks, they don't focus on converting UI images to HTML code. To address this, we propose \ours, a fine-tuning method that equips MLLMs with domain-specific knowledge needed for UI-to-HTML generation.

\subsection{Attention Mechanism}
\noindent The attention mechanism is the key part of modern Transformer~\cite{transformer} architectures, as it effectively captures the hidden features of input data. To handle the challenges of certain domains, specialized attention mechanisms have been explored, such as pyramid attention~\cite{pyramid} and hierarchical attention~\cite{longcoder, cast, hierarchynet, hierarchical} designed for long-range, cross-file code understanding and generation, as well as regularized attention for assembly code~\cite{codeart}. Different from existing work, \ours targets HTML code, with the new challenge of its restricted structure. \ours{} designs a novel \attn{} to learn such structure knowledge.

\subsection{UI to HTML Generation}
\noindent Early direction for UI code generation utilizes sketch webpage figures, e.g., hand-drawn website sketches, to generate UI code that can be rendered into similar images as sketch images~\cite{related_background3_sketch2code}. Yet, this direction is not practical, as not all front-end developers want to draw out a sketch website when they need help from an automated tool. In contrast, leveraging advancements in MLLMs, Huggingface has recently released WebSight, which is trained on the WebSight-v0.1 dataset~\cite{related1websight}. Although specific details of the model are not disclosed, it represents a significant shift towards end-to-end UI to code generation. Similarly, Design2Code-18B is a model using CogAgent as the backbone using a subset of WebSight-v0.1~\cite{related3design2code, cogagent}. However, neither WebSight nor Design2Code tries to adapt domain knowledge of HTML for this task. In contrast, we provide structure-aware attention and apply contrastive learning with the mutations to teach the model the fine-grained difference of HTML images.

\section{Conclusion}
\noindent This work presents \ours{}, a fine-tuning pipeline for UI-to-HTML code generation, that is generalizable to any transformer-based MLLMs. \ours{} introduces a structure-aware attention mechanism to capture HTML structure and employs contrastive learning to align visual and textual understanding, aiding MLLMs in distinguishing subtle webpage differences. \ours{} outperforms standard fine-tuning on two backbone MLLMs, with improvements of up to 9 pp in HTML Match, 0.0982 in CW-SSIM, 32.99 in CLIP, and 27.12 pp in LLEM. Ablation studies confirm that both key components of \ours{} contribute to better cross-modality understanding and more robust code generation. Notably, \ours{} is model-independent and can enhance any MLLMs for UI-to-HTML code generation.

\section{Limitation}
One limitation of \ours{} is that it has only been implemented on two models: VLM-WebSight and Moondream2. While \ours{} could potentially be applied to any MLLM, like Design2Code, further exploration is limited by computing resources. Our experiments show that \ours{} brings significant improvements over standard fine-tuning on these two models, indicating some level of generalizability.
Another limitation is that the metrics we use do not fully capture human evaluation. HTML-Match overlooks CSS styling, and metrics like CW-SSIM, CLIP, and LLEM are similarity-based, which can lead to unreliable scores. Evaluating HTML code automatically is challenging, so we include CLIP and LLEM, as used in previous work~\cite{related3design2code, related5vision2ui}, along with CW-SSIM and HTML-Match to ensure fair evaluation.

\section{Acknowledgements}
We thank the anonymous reviewers for their feedback on
this work. This research was supported in part by NSF 1901242 and 2006688 and a CFI fund. This work also used Anvil at Purdue University through allocation CIS240304 from the Advanced Cyberinfrastructure Coordination Ecosystem: Services \& Support (ACCESS) program~\cite{ack-access}, which is supported by U.S. National Science Foundation grants \#2138259, \#2138286, \#2138307, \#2137603, and \#2138296.

\bibliography{main}

\begin{thebibliography}{43}
\providecommand{\natexlab}[1]{#1}

\bibitem[{Bai et~al.(2023)Bai, Bai, Yang, Wang, Tan, Wang, Lin, Zhou, and
  Zhou}]{Qwen-VL}
Jinze Bai, Shuai Bai, Shusheng Yang, Shijie Wang, Sinan Tan, Peng Wang, Junyang
  Lin, Chang Zhou, and Jingren Zhou. 2023.
\newblock Qwen-vl: A frontier large vision-language model with versatile
  abilities.
\newblock \emph{arXiv preprint arXiv:2308.12966}.

\bibitem[{Blecher et~al.(2023)Blecher, Cucurull, Scialom, and Stojnic}]{nougat}
Lukas Blecher, Guillem Cucurull, Thomas Scialom, and Robert Stojnic. 2023.
\newblock \href {https://arxiv.org/abs/2308.13418} {Nougat: Neural optical
  understanding for academic documents}.
\newblock \emph{Preprint}, arXiv:2308.13418.

\bibitem[{Boerner et~al.(2023)Boerner, Deems, Furlani, Knuth, and
  Towns}]{ack-access}
Timothy~J. Boerner, Stephen Deems, Thomas~R. Furlani, Shelley~L. Knuth, and
  John Towns. 2023.
\newblock \href {https://doi.org/10.1145/3569951.3597559} {Access: Advancing
  innovation: Nsf’s advanced cyberinfrastructure coordination ecosystem:
  Services \& support}.
\newblock PEARC '23, page 173–176, New York, NY, USA. Association for
  Computing Machinery.

\bibitem[{Chai and Li(2022)}]{pyramid}
Lei Chai and Ming Li. 2022.
\newblock Pyramid attention for source code summarization.
\newblock \emph{Advances in Neural Information Processing Systems},
  35:20421--20433.

\bibitem[{Chen et~al.(2023)Chen, Li, Dong, Zhang, He, Wang, Zhao, and
  Lin}]{sharegpt4v}
Lin Chen, Jisong Li, Xiaoyi Dong, Pan Zhang, Conghui He, Jiaqi Wang, Feng Zhao,
  and Dahua Lin. 2023.
\newblock Sharegpt4v: Improving large multi-modal models with better captions.
\newblock \emph{arXiv preprint arXiv:2311.12793}.

\bibitem[{Dai et~al.(2023)Dai, Li, Li, Tiong, Zhao, Wang, Li, Fung, and
  Hoi}]{instructblip}
Wenliang Dai, Junnan Li, Dongxu Li, Anthony Meng~Huat Tiong, Junqi Zhao,
  Weisheng Wang, Boyang Li, Pascale Fung, and Steven Hoi. 2023.
\newblock \href {https://arxiv.org/abs/2305.06500} {Instructblip: Towards
  general-purpose vision-language models with instruction tuning}.
\newblock \emph{Preprint}, arXiv:2305.06500.

\bibitem[{Fried et~al.(2023)Fried, Aghajanyan, Lin, Wang, Wallace, Shi, Zhong,
  tau Yih, Zettlemoyer, and Lewis}]{incoder}
Daniel Fried, Armen Aghajanyan, Jessy Lin, Sida Wang, Eric Wallace, Freda Shi,
  Ruiqi Zhong, Wen tau Yih, Luke Zettlemoyer, and Mike Lewis. 2023.
\newblock \href {https://arxiv.org/abs/2204.05999} {Incoder: A generative model
  for code infilling and synthesis}.
\newblock \emph{Preprint}, arXiv:2204.05999.

\bibitem[{Gao et~al.(2021)Gao, Yao, and Chen}]{related9constrastive-simcse}
Tianyu Gao, Xingcheng Yao, and Danqi Chen. 2021.
\newblock \href {https://doi.org/10.18653/v1/2021.emnlp-main.552} {{S}im{CSE}:
  Simple contrastive learning of sentence embeddings}.
\newblock In \emph{Proceedings of the 2021 Conference on Empirical Methods in
  Natural Language Processing}, pages 6894--6910, Online and Punta Cana,
  Dominican Republic. Association for Computational Linguistics.

\bibitem[{Gui et~al.(2024)Gui, Li, Wan, Shi, Zhang, Su, Dong, Zhou, and
  Jiang}]{related5vision2ui}
Yi~Gui, Zhen Li, Yao Wan, Yemin Shi, Hongyu Zhang, Yi~Su, Shaoling Dong, Xing
  Zhou, and Wenbin Jiang. 2024.
\newblock \href {https://arxiv.org/abs/2404.06369} {Vision2ui: A real-world
  dataset with layout for code generation from ui designs}.
\newblock \emph{Preprint}, arXiv:2404.06369.

\bibitem[{Guo et~al.(2023)Guo, Xu, Duan, Yin, and McAuley}]{longcoder}
Daya Guo, Canwen Xu, Nan Duan, Jian Yin, and Julian McAuley. 2023.
\newblock Longcoder: a long-range pre-trained language model for code
  completion.
\newblock In \emph{Proceedings of the 40th International Conference on Machine
  Learning}, ICML'23. JMLR.org.

\bibitem[{Guo et~al.(2024)Guo, Zhu, Yang, Xie, Dong, Zhang, Chen, Bi, Wu, Li,
  Luo, Xiong, and Liang}]{related8deepseekcoder}
Daya Guo, Qihao Zhu, Dejian Yang, Zhenda Xie, Kai Dong, Wentao Zhang, Guanting
  Chen, Xiao Bi, Y.~Wu, Y.~K. Li, Fuli Luo, Yingfei Xiong, and Wenfeng Liang.
  2024.
\newblock \href {https://arxiv.org/abs/2401.14196} {Deepseek-coder: When the
  large language model meets programming -- the rise of code intelligence}.
\newblock \emph{Preprint}, arXiv:2401.14196.

\bibitem[{Hong et~al.(2023)Hong, Wang, Lv, Xu, Yu, Ji, Wang, Wang, Zhang, Li,
  Xu, Dong, Ding, and Tang}]{cogagent}
Wenyi Hong, Weihan Wang, Qingsong Lv, Jiazheng Xu, Wenmeng Yu, Junhui Ji, Yan
  Wang, Zihan Wang, Yuxuan Zhang, Juanzi Li, Bin Xu, Yuxiao Dong, Ming Ding,
  and Jie Tang. 2023.
\newblock \href {https://arxiv.org/abs/2312.08914} {Cogagent: A visual language
  model for gui agents}.
\newblock \emph{Preprint}, arXiv:2312.08914.

\bibitem[{Hu et~al.(2022)Hu, Shen, Wallis, Allen-Zhu, Li, Wang, Wang, and
  Chen}]{lora}
Edward~J Hu, Yelong Shen, Phillip Wallis, Zeyuan Allen-Zhu, Yuanzhi Li, Shean
  Wang, Lu~Wang, and Weizhu Chen. 2022.
\newblock \href {https://openreview.net/forum?id=nZeVKeeFYf9} {Lo{RA}: Low-rank
  adaptation of large language models}.
\newblock In \emph{International Conference on Learning Representations}.

\bibitem[{Jiang et~al.(2023)Jiang, Sablayrolles, Mensch, Bamford, Chaplot,
  de~las Casas, Bressand, Lengyel, Lample, Saulnier, Lavaud, Lachaux, Stock,
  Scao, Lavril, Wang, Lacroix, and Sayed}]{mistral}
Albert~Q. Jiang, Alexandre Sablayrolles, Arthur Mensch, Chris Bamford,
  Devendra~Singh Chaplot, Diego de~las Casas, Florian Bressand, Gianna Lengyel,
  Guillaume Lample, Lucile Saulnier, Lélio~Renard Lavaud, Marie-Anne Lachaux,
  Pierre Stock, Teven~Le Scao, Thibaut Lavril, Thomas Wang, Timothée Lacroix,
  and William~El Sayed. 2023.
\newblock \href {https://arxiv.org/abs/2310.06825} {Mistral 7b}.
\newblock \emph{Preprint}, arXiv:2310.06825.

\bibitem[{Laurençon et~al.(2023)Laurençon, Saulnier, Tronchon, Bekman, Singh,
  Lozhkov, Wang, Karamcheti, Rush, Kiela, Cord, and Sanh}]{idefmics}
Hugo Laurençon, Lucile Saulnier, Léo Tronchon, Stas Bekman, Amanpreet Singh,
  Anton Lozhkov, Thomas Wang, Siddharth Karamcheti, Alexander~M. Rush, Douwe
  Kiela, Matthieu Cord, and Victor Sanh. 2023.
\newblock \href {https://arxiv.org/abs/2306.16527} {Obelics: An open web-scale
  filtered dataset of interleaved image-text documents}.
\newblock \emph{Preprint}, arXiv:2306.16527.

\bibitem[{Laurençon et~al.(2024)Laurençon, Tronchon, and
  Sanh}]{related1websight}
Hugo Laurençon, Léo Tronchon, and Victor Sanh. 2024.
\newblock \href {https://arxiv.org/abs/2403.09029} {Unlocking the conversion of
  web screenshots into html code with the websight dataset}.
\newblock \emph{Preprint}, arXiv:2403.09029.

\bibitem[{Li et~al.(2023{\natexlab{a}})Li, Li, Le, Wang, Savarese, and
  Hoi}]{lavis}
Dongxu Li, Junnan Li, Hung Le, Guangsen Wang, Silvio Savarese, and Steven~C.H.
  Hoi. 2023{\natexlab{a}}.
\newblock \href {https://aclanthology.org/2023.acl-demo.3} {{LAVIS}: A one-stop
  library for language-vision intelligence}.
\newblock In \emph{Proceedings of the 61st Annual Meeting of the Association
  for Computational Linguistics (Volume 3: System Demonstrations)}, pages
  31--41, Toronto, Canada. Association for Computational Linguistics.

\bibitem[{Li et~al.(2023{\natexlab{b}})Li, Li, Savarese, and Hoi}]{blip2}
Junnan Li, Dongxu Li, Silvio Savarese, and Steven Hoi. 2023{\natexlab{b}}.
\newblock Blip-2: bootstrapping language-image pre-training with frozen image
  encoders and large language models.
\newblock In \emph{Proceedings of the 40th International Conference on Machine
  Learning}, ICML'23. JMLR.org.

\bibitem[{Li et~al.(2022)Li, Li, Xiong, and Hoi}]{blip}
Junnan Li, Dongxu Li, Caiming Xiong, and Steven Hoi. 2022.
\newblock \href {https://proceedings.mlr.press/v162/li22n.html} {{BLIP}:
  Bootstrapping language-image pre-training for unified vision-language
  understanding and generation}.
\newblock In \emph{Proceedings of the 39th International Conference on Machine
  Learning}, volume 162 of \emph{Proceedings of Machine Learning Research},
  pages 12888--12900. PMLR.

\bibitem[{Li et~al.(2023{\natexlab{c}})Li, Allal, Zi, Muennighoff, Kocetkov,
  Mou, Marone, Akiki, Li, Chim, Liu, Zheltonozhskii, Zhuo, Wang, Dehaene,
  Davaadorj, Lamy-Poirier, Monteiro, Shliazhko, Gontier, Meade, Zebaze, Yee,
  Umapathi, Zhu, Lipkin, Oblokulov, Wang, Murthy, Stillerman, Patel,
  Abulkhanov, Zocca, Dey, Zhang, Fahmy, Bhattacharyya, Yu, Singh, Luccioni,
  Villegas, Kunakov, Zhdanov, Romero, Lee, Timor, Ding, Schlesinger,
  Schoelkopf, Ebert, Dao, Mishra, Gu, Robinson, Anderson, Dolan-Gavitt,
  Contractor, Reddy, Fried, Bahdanau, Jernite, Ferrandis, Hughes, Wolf, Guha,
  von Werra, and de~Vries}]{starcoder}
Raymond Li, Loubna~Ben Allal, Yangtian Zi, Niklas Muennighoff, Denis Kocetkov,
  Chenghao Mou, Marc Marone, Christopher Akiki, Jia Li, Jenny Chim, Qian Liu,
  Evgenii Zheltonozhskii, Terry~Yue Zhuo, Thomas Wang, Olivier Dehaene, Mishig
  Davaadorj, Joel Lamy-Poirier, João Monteiro, Oleh Shliazhko, Nicolas
  Gontier, Nicholas Meade, Armel Zebaze, Ming-Ho Yee, Logesh~Kumar Umapathi,
  Jian Zhu, Benjamin Lipkin, Muhtasham Oblokulov, Zhiruo Wang, Rudra Murthy,
  Jason Stillerman, Siva~Sankalp Patel, Dmitry Abulkhanov, Marco Zocca, Manan
  Dey, Zhihan Zhang, Nour Fahmy, Urvashi Bhattacharyya, Wenhao Yu, Swayam
  Singh, Sasha Luccioni, Paulo Villegas, Maxim Kunakov, Fedor Zhdanov, Manuel
  Romero, Tony Lee, Nadav Timor, Jennifer Ding, Claire Schlesinger, Hailey
  Schoelkopf, Jan Ebert, Tri Dao, Mayank Mishra, Alex Gu, Jennifer Robinson,
  Carolyn~Jane Anderson, Brendan Dolan-Gavitt, Danish Contractor, Siva Reddy,
  Daniel Fried, Dzmitry Bahdanau, Yacine Jernite, Carlos~Muñoz Ferrandis, Sean
  Hughes, Thomas Wolf, Arjun Guha, Leandro von Werra, and Harm de~Vries.
  2023{\natexlab{c}}.
\newblock \href {https://arxiv.org/abs/2305.06161} {Starcoder: may the source
  be with you!}
\newblock \emph{Preprint}, arXiv:2305.06161.

\bibitem[{Liu et~al.(2023{\natexlab{a}})Liu, Li, Li, and Lee}]{llava1.5}
Haotian Liu, Chunyuan Li, Yuheng Li, and Yong~Jae Lee. 2023{\natexlab{a}}.
\newblock Improved baselines with visual instruction tuning.

\bibitem[{Liu et~al.(2023{\natexlab{b}})Liu, Li, Wu, and Lee}]{llava}
Haotian Liu, Chunyuan Li, Qingyang Wu, and Yong~Jae Lee. 2023{\natexlab{b}}.
\newblock \href {https://arxiv.org/abs/2304.08485} {Visual instruction tuning}.
\newblock \emph{Preprint}, arXiv:2304.08485.

\bibitem[{Liu et~al.(2024)Liu, Wang, Yin, Molchanov, Wang, Cheng, and
  Chen}]{dora}
Shih-Yang Liu, Chien-Yi Wang, Hongxu Yin, Pavlo Molchanov, Yu-Chiang~Frank
  Wang, Kwang-Ting Cheng, and Min-Hung Chen. 2024.
\newblock \href {arxiv.org/abs/2402.09353} {{DoRA}: Weight-decomposed low-rank
  adaptation}.

\bibitem[{Loshchilov and Hutter(2019)}]{adamw}
Ilya Loshchilov and Frank Hutter. 2019.
\newblock \href {https://arxiv.org/abs/1711.05101} {Decoupled weight decay
  regularization}.
\newblock \emph{Preprint}, arXiv:1711.05101.

\bibitem[{Lozhkov et~al.(2024)Lozhkov, Li, Allal, Cassano, Lamy-Poirier, Tazi,
  Tang, Pykhtar, Liu, Wei, Liu, Tian, Kocetkov, Zucker, Belkada, Wang, Liu,
  Abulkhanov, Paul, Li, Li, Risdal, Li, Zhu, Zhuo, Zheltonozhskii, Dade, Yu,
  Krauß, Jain, Su, He, Dey, Abati, Chai, Muennighoff, Tang, Oblokulov, Akiki,
  Marone, Mou, Mishra, Gu, Hui, Dao, Zebaze, Dehaene, Patry, Xu, McAuley, Hu,
  Scholak, Paquet, Robinson, Anderson, Chapados, Patwary, Tajbakhsh, Jernite,
  Ferrandis, Zhang, Hughes, Wolf, Guha, von Werra, and de~Vries}]{starcoder2}
Anton Lozhkov, Raymond Li, Loubna~Ben Allal, Federico Cassano, Joel
  Lamy-Poirier, Nouamane Tazi, Ao~Tang, Dmytro Pykhtar, Jiawei Liu, Yuxiang
  Wei, Tianyang Liu, Max Tian, Denis Kocetkov, Arthur Zucker, Younes Belkada,
  Zijian Wang, Qian Liu, Dmitry Abulkhanov, Indraneil Paul, Zhuang Li, Wen-Ding
  Li, Megan Risdal, Jia Li, Jian Zhu, Terry~Yue Zhuo, Evgenii Zheltonozhskii,
  Nii Osae~Osae Dade, Wenhao Yu, Lucas Krauß, Naman Jain, Yixuan Su, Xuanli
  He, Manan Dey, Edoardo Abati, Yekun Chai, Niklas Muennighoff, Xiangru Tang,
  Muhtasham Oblokulov, Christopher Akiki, Marc Marone, Chenghao Mou, Mayank
  Mishra, Alex Gu, Binyuan Hui, Tri Dao, Armel Zebaze, Olivier Dehaene, Nicolas
  Patry, Canwen Xu, Julian McAuley, Han Hu, Torsten Scholak, Sebastien Paquet,
  Jennifer Robinson, Carolyn~Jane Anderson, Nicolas Chapados, Mostofa Patwary,
  Nima Tajbakhsh, Yacine Jernite, Carlos~Muñoz Ferrandis, Lingming Zhang, Sean
  Hughes, Thomas Wolf, Arjun Guha, Leandro von Werra, and Harm de~Vries. 2024.
\newblock \href {https://arxiv.org/abs/2402.19173} {Starcoder 2 and the stack
  v2: The next generation}.
\newblock \emph{Preprint}, arXiv:2402.19173.

\bibitem[{Nguyen et~al.(2023)Nguyen, Bui, Hy, Tran-Thanh, and
  Nguyen}]{hierarchynet}
Minh~Huynh Nguyen, Nghi D.~Q. Bui, Truong~Son Hy, Long Tran-Thanh, and Tien~N.
  Nguyen. 2023.
\newblock \href {https://arxiv.org/abs/2205.15479} {Hierarchynet: Learning to
  summarize source code with heterogeneous representations}.
\newblock \emph{Preprint}, arXiv:2205.15479.

\bibitem[{Nijkamp et~al.(2022)Nijkamp, Pang, Hayashi, Tu, Wang, Zhou, Savarese,
  and Xiong}]{codegen}
Erik Nijkamp, Bo~Pang, Hiroaki Hayashi, Lifu Tu, Huan Wang, Yingbo Zhou, Silvio
  Savarese, and Caiming Xiong. 2022.
\newblock A conversational paradigm for program synthesis.
\newblock \emph{arXiv preprint}.

\bibitem[{Radford et~al.(2021)Radford, Kim, Hallacy, Ramesh, Goh, Agarwal,
  Sastry, Askell, Mishkin, Clark, Krueger, and Sutskever}]{related6clip}
Alec Radford, Jong~Wook Kim, Chris Hallacy, Aditya Ramesh, Gabriel Goh,
  Sandhini Agarwal, Girish Sastry, Amanda Askell, Pamela Mishkin, Jack Clark,
  Gretchen Krueger, and Ilya Sutskever. 2021.
\newblock \href {https://api.semanticscholar.org/CorpusID:231591445} {Learning
  transferable visual models from natural language supervision}.
\newblock In \emph{International Conference on Machine Learning}.

\bibitem[{Ramesh et~al.(2022)Ramesh, Dhariwal, Nichol, Chu, and
  Chen}]{related-dalle2}
Aditya Ramesh, Prafulla Dhariwal, Alex Nichol, Casey Chu, and Mark Chen. 2022.
\newblock \href {https://arxiv.org/abs/2204.06125} {Hierarchical
  text-conditional image generation with clip latents}.
\newblock \emph{Preprint}, arXiv:2204.06125.

\bibitem[{Robinson(2019)}]{related_background3_sketch2code}
Alex Robinson. 2019.
\newblock \href {https://arxiv.org/abs/1905.13750} {Sketch2code: Generating a
  website from a paper mockup}.
\newblock \emph{Preprint}, arXiv:1905.13750.

\bibitem[{Rombach et~al.(2021)Rombach, Blattmann, Lorenz, Esser, and
  Ommer}]{related-stablediffusion}
Robin Rombach, A.~Blattmann, Dominik Lorenz, Patrick Esser, and Bj{\"o}rn
  Ommer. 2021.
\newblock \href {https://api.semanticscholar.org/CorpusID:245335280}
  {High-resolution image synthesis with latent diffusion models}.
\newblock \emph{2022 IEEE/CVF Conference on Computer Vision and Pattern
  Recognition (CVPR)}, pages 10674--10685.

\bibitem[{Rozière et~al.(2024)Rozière, Gehring, Gloeckle, Sootla, Gat, Tan,
  Adi, Liu, Sauvestre, Remez, Rapin, Kozhevnikov, Evtimov, Bitton, Bhatt,
  Ferrer, Grattafiori, Xiong, Défossez, Copet, Azhar, Touvron, Martin,
  Usunier, Scialom, and Synnaeve}]{codellama}
Baptiste Rozière, Jonas Gehring, Fabian Gloeckle, Sten Sootla, Itai Gat,
  Xiaoqing~Ellen Tan, Yossi Adi, Jingyu Liu, Romain Sauvestre, Tal Remez,
  Jérémy Rapin, Artyom Kozhevnikov, Ivan Evtimov, Joanna Bitton, Manish
  Bhatt, Cristian~Canton Ferrer, Aaron Grattafiori, Wenhan Xiong, Alexandre
  Défossez, Jade Copet, Faisal Azhar, Hugo Touvron, Louis Martin, Nicolas
  Usunier, Thomas Scialom, and Gabriel Synnaeve. 2024.
\newblock \href {https://arxiv.org/abs/2308.12950} {Code llama: Open foundation
  models for code}.
\newblock \emph{Preprint}, arXiv:2308.12950.

\bibitem[{Sampat et~al.(2009)Sampat, Wang, Gupta, Bovik, and
  Markey}]{metrics-cwssim}
Mehul~P. Sampat, Zhou Wang, Shalini Gupta, Alan~Conrad Bovik, and Mia~K.
  Markey. 2009.
\newblock \href {https://doi.org/10.1109/TIP.2009.2025923} {Complex wavelet
  structural similarity: A new image similarity index}.
\newblock \emph{IEEE Transactions on Image Processing}, 18(11):2385--2401.

\bibitem[{Shi et~al.(2021)Shi, Wang, Du, Zhang, Han, Zhang, and Sun}]{cast}
Ensheng Shi, Yanlin Wang, Lun Du, Hongyu Zhang, Shi Han, Dongmei Zhang, and
  Hongbin Sun. 2021.
\newblock \href {https://doi.org/10.18653/v1/2021.emnlp-main.332} {{CAST}:
  Enhancing code summarization with hierarchical splitting and reconstruction
  of abstract syntax trees}.
\newblock In \emph{Proceedings of the 2021 Conference on Empirical Methods in
  Natural Language Processing}, pages 4053--4062, Online and Punta Cana,
  Dominican Republic. Association for Computational Linguistics.

\bibitem[{Si et~al.(2024)Si, Zhang, Yang, Liu, and Yang}]{related3design2code}
Chenglei Si, Yanzhe Zhang, Zhengyuan Yang, Ruibo Liu, and Diyi Yang. 2024.
\newblock \href {https://arxiv.org/abs/2403.03163} {Design2code: How far are we
  from automating front-end engineering?}
\newblock \emph{Preprint}, arXiv:2403.03163.

\bibitem[{Su et~al.(2024)Su, Xu, Huang, Zhang, Ye, Huang, and Zhang}]{codeart}
Zian Su, Xiangzhe Xu, Ziyang Huang, Zhuo Zhang, Yapeng Ye, Jianjun Huang, and
  Xiangyu Zhang. 2024.
\newblock \href {https://doi.org/10.1145/3643752} {Codeart: Better code models
  by attention regularization when symbols are lacking}.
\newblock \emph{Proc. ACM Softw. Eng.}, 1(FSE).

\bibitem[{Suri et~al.(2024)Suri, Mathur, Dernoncourt, Jain, Morariu, Sawhney,
  Nakov, and Manocha}]{related-docedit}
Manan Suri, Puneet Mathur, Franck Dernoncourt, Rajiv Jain, Vlad~I Morariu,
  Ramit Sawhney, Preslav Nakov, and Dinesh Manocha. 2024.
\newblock \href {https://doi.org/10.18653/v1/2024.emnlp-main.867}
  {{D}oc{E}dit-v2: Document structure editing via multimodal {LLM} grounding}.
\newblock In \emph{Proceedings of the 2024 Conference on Empirical Methods in
  Natural Language Processing}, pages 15485--15505, Miami, Florida, USA.
  Association for Computational Linguistics.

\bibitem[{Touvron et~al.(2023)Touvron, Lavril, Izacard, Martinet, Lachaux,
  Lacroix, Rozière, Goyal, Hambro, Azhar, Rodriguez, Joulin, Grave, and
  Lample}]{related_background2_llama}
Hugo Touvron, Thibaut Lavril, Gautier Izacard, Xavier Martinet, Marie-Anne
  Lachaux, Timothée Lacroix, Baptiste Rozière, Naman Goyal, Eric Hambro,
  Faisal Azhar, Aurelien Rodriguez, Armand Joulin, Edouard Grave, and Guillaume
  Lample. 2023.
\newblock \href {https://arxiv.org/abs/2302.13971} {Llama: Open and efficient
  foundation language models}.
\newblock \emph{Preprint}, arXiv:2302.13971.

\bibitem[{Vaswani et~al.(2017)Vaswani, Shazeer, Parmar, Uszkoreit, Jones,
  Gomez, Kaiser, and Polosukhin}]{transformer}
Ashish Vaswani, Noam Shazeer, Niki Parmar, Jakob Uszkoreit, Llion Jones,
  Aidan~N Gomez, \L~ukasz Kaiser, and Illia Polosukhin. 2017.
\newblock \href
  {https://proceedings.neurips.cc/paper_files/paper/2017/file/3f5ee243547dee91fbd053c1c4a845aa-Paper.pdf}
  {Attention is all you need}.
\newblock In \emph{Advances in Neural Information Processing Systems},
  volume~30. Curran Associates, Inc.

\bibitem[{vikhyat(2024)}]{moondream}
vikhyat. 2024.
\newblock \href {https://github.com/vikhyat/moondream} {Moondream: tiny vision
  language model}.

\bibitem[{Wei et~al.(2023)Wei, Kong, Chen, Zhao, Ge, Yang, Sun, Han, and
  Zhang}]{vary}
Haoran Wei, Lingyu Kong, Jinyue Chen, Liang Zhao, Zheng Ge, Jinrong Yang,
  Jianjian Sun, Chunrui Han, and Xiangyu Zhang. 2023.
\newblock Vary: Scaling up the vision vocabulary for large vision-language
  models.
\newblock \emph{arXiv preprint arXiv:2312.06109}.

\bibitem[{Yang et~al.(2016)Yang, Yang, Dyer, He, Smola, and
  Hovy}]{hierarchical}
Zichao Yang, Diyi Yang, Chris Dyer, Xiaodong He, Alex Smola, and Eduard Hovy.
  2016.
\newblock \href {https://doi.org/10.18653/v1/N16-1174} {Hierarchical attention
  networks for document classification}.
\newblock In \emph{Proceedings of the 2016 Conference of the North {A}merican
  Chapter of the Association for Computational Linguistics: Human Language
  Technologies}, pages 1480--1489, San Diego, California. Association for
  Computational Linguistics.

\bibitem[{Zhai et~al.(2023)Zhai, Mustafa, Kolesnikov, and
  Beyer}]{related7siglip}
X.~Zhai, B.~Mustafa, A.~Kolesnikov, and L.~Beyer. 2023.
\newblock \href {https://doi.org/10.1109/ICCV51070.2023.01100} {Sigmoid loss
  for language image pre-training}.
\newblock In \emph{2023 IEEE/CVF International Conference on Computer Vision
  (ICCV)}, pages 11941--11952, Los Alamitos, CA, USA. IEEE Computer Society.

\end{thebibliography}

\appendix
\section{Appendix}

\subsection{Mutation Rules}
\label{sec: appendix_mutation_rule}

Table~\ref{table:mutation_specification} shows the mutation rules we used to mutate the HTML code and create the contrastive learning dataset.
Both HTML code and the CSS styles for each element are mutated according to the failure types in our manual analysis

For CSS styles, we mutate the properties of 1) color, 2) size, 3) margin, 4) font size, 5) type of the element bounding box (display), and 6) positioning of each element. Column ``Specification'' in \Cref{table:mutation_specification} shows the details of the valid values for each property. For HTML codes, we randomly duplicate one HTML element excluding the ones that will cause render failures (i.e., \code{<head>}, \code{<header>}, \code{<html>}, \code{<body>}).

\begin{table}[h]
\centering
\scriptsize
\setlength{\tabcolsep}{2.1pt}
    \begin{tabular}{lll}
    \toprule
    Class & Failure Type & Specification \\
    \midrule
    \multirow{8}{*}{CSS} & Color & Random Color in Range [\code{\#000000, \#FFFFFF}] \\
    & \cellcolor{lightgray}Size & \cellcolor{lightgray}Random Size in [0, 500] pixels \\
    & Margin & Random Size in [0, 100] pixels \\
    & \cellcolor{lightgray}Font  & \cellcolor{lightgray}Random Size in [0, 40] pixels \\
    & Display & Random Keyword for \code{text-align}, \code{display}, \\
    &  & \code{flex-direction}, and \code{justify-content} \\
    & \cellcolor{lightgray}Position & \cellcolor{lightgray}Random Keyword for \code{border-radius}, \\
    & \cellcolor{lightgray} & \cellcolor{lightgray}\code{position}, \code{top}, and \code{right} \\
        \midrule
    \multirow{2}{*}{HTML} & \multirow{2}{*}{Structure} & Duplication of a Random HTML Element, excluding \\
    & & \code{<head>}, \code{<header>}, \code{<html>}, \code{<body>} \\
    \bottomrule
    \end{tabular}
\caption{Specification for Mutation Rules to construct the Contrastive dataset.}
\label{table:mutation_specification}
\end{table}

\subsection{Illustrating Example}
\label{challeng2_example}
\begin{figure}[t]
    \centering
    \includegraphics[width=\linewidth]{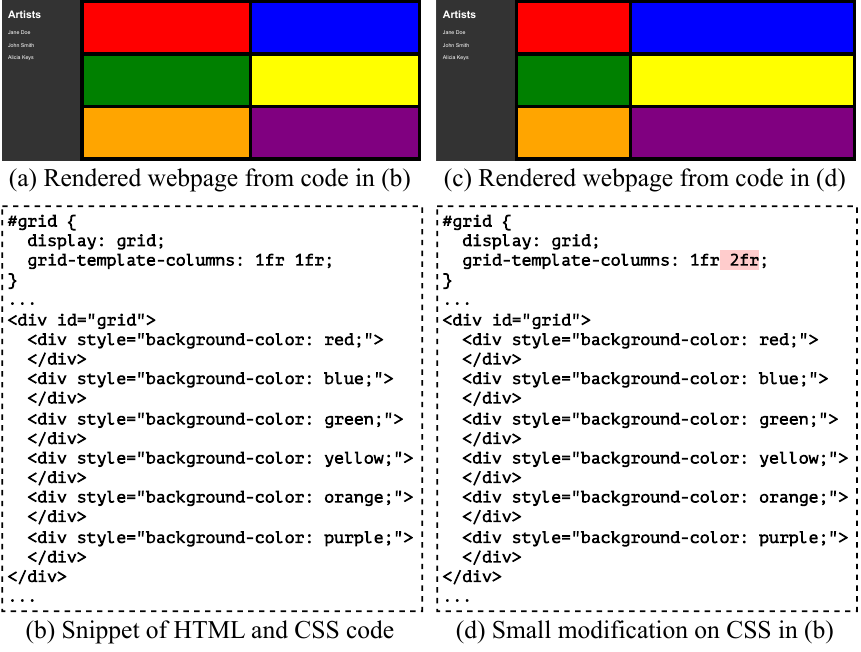}
    \caption{
    Existing MLLM generates identical HTML and CSS  code in (b) given the different webpage screenshots in (a) and (c). The MLLM fails to generate the correct \colorbox{lightred}{\code{2fr}} highlighted in red in (d).}
    \label{fig:contrastive-motivation}
\end{figure}

\Cref{fig:contrastive-motivation} illustrates the second challenge, i.e., learning the fine differences and details of the visual input. 
\Cref{fig:contrastive-motivation}(a) and \Cref{fig:contrastive-motivation}(c) are two highly similar but different UI images of rendered webpages, i.e., the colors and text are identical, but the widths of the columns are slightly different. 
VLM-WebSight~\cite{related1websight}, a state-of-the-art MLLM for webpage image to HTML code generation, fails to capture such small differences; thus, it generates identical HTML and CSS code (\Cref{fig:contrastive-motivation}(b)) for the different UI images: \Cref{fig:contrastive-motivation}(a) and \Cref{fig:contrastive-motivation}(c). 
The model fails to generate \code{1fr \colorbox{lightred}{2fr}} (highlighted in code segments (d) with red background) for screenshot (c).
In this case, VLM-WebSight's vision model fails to recognize the visual difference, and its text model is unable to use the encoded visual information to produce accurate textual output.

\subsection{Tuning the Integration of \ours{}'s Structure-Aware Attention.} 
\ours{} applies the \attn{} on the attention layer in MLLM's decoder. To study the portion of attention heads that use \attn{},
we fine-tuned VLM-WebSight on a subset of our training dataset (40,000 pairs of HTML code and webpage screenshots). We set the portion of attention heads using \attn{} from $\frac{1}{8}$ to all, incrementing each setting by $\frac{1}{8}$. All models are trained with a batch size of 4, and a learning rate of $2e^{-5}$. The models are then evaluated on the validation dataset. We use two metrics to decide the final hyper-parameter: averaged LLEM score and training loss. 

\begin{figure}[t]
    \centering
    \includegraphics[width=\linewidth]{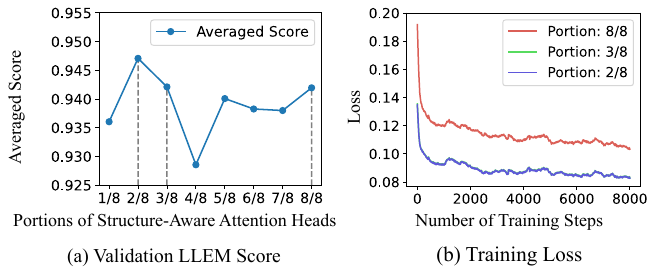}
    \caption{Illustration of the tuning process of the parameter that controls the effect of structure-aware attention. In (b), the green line almost overlaps with the blue line.}
    \label{fig:attention_head}
\end{figure}

\Cref{fig:attention_head} (a) shows the averaged LLEM score. Applying \attn{} on $\frac{2}{8}$, $\frac{3}{8}$, and $\frac{8}{8}$ of the attention heads results in the top three validation scores. We also consider their training loss in~\Cref{fig:attention_head} (b). Although applying \attn{} on all (i.e., $\frac{8}{8}$) of the attention heads yields a high LLEM score, it also results in a high training loss, likely due to the regular attention heads retaining some prior knowledge during pre-training. In contrast, $\frac{2}{8}$, $\frac{3}{8}$ show similar and lower training losses. Combining these results, we select $\frac{2}{8}$ (i.e., $\frac{1}{4}$) as the final hyper-parameter controlling the portion of attention heads that use \attn{}.

\subsection{Additional Ablation of Contrastive Learning's effect}

\begin{table}[t]
\centering
\scriptsize
\setlength{\tabcolsep}{8pt}
\begin{tabular}{lccc}
\toprule
   {Techniques} & $d(\overline{v^i}, c) \uparrow$ & $sim(\overline{v^i}, c_g) \downarrow$ \\
\midrule
Standard FT &  0.1224 & 0.9910 \\
\noattn &  0.7590 & 0.6202 \\
\bottomrule
\end{tabular}
\caption{Distance ($d$) and similarity ($sim$) between each averaged image embeddings $\overline{v^i}$ with the corresponding centroid $c$ of the group of mutants, with Moondream2 backbone.}
\label{table:rq4_image2image}
\end{table}

\newsubsubsection{Capturing subtle visual differences.}
Using the same computed embeddings, we compute the averaged distances and similarities between each image embedding the centroid of its corresponding group of mutants. Formally, for each group of mutants, $G$, consisting of image embedding $\{\overline{v^i}\},\ v^i\in G$, the centroid of the image embeddings is computed as:  
$c = \sum^{|G|}_{i = 1} \overline{v^i}/ {|G|}$.
\Cref{table:rq4_image2image} shows the distance and cosine similarities between the image embeddings. The average distance between each image embedding with its respective centroid computed by the \noattn is 0.7590, greatly surpassing the average distance of 0.1224 computed by Standard FT. Likewise, the cosine similarity between the image embeddings is much lower for \noattn (0.6202 vs. 0.9910), showing \noattn{'s} better ability to distinguish between the images. 

\Cref{fig:case_study_text_and_image} also shows that Standard FT encodes the four different images almost the same in the latent space (i.e., the four red circles are overlapped in (a)), while \noattn is able to encode them differently.

\subsection{Attention Visualization}
\begin{figure*}[t]
    \centering
    \includegraphics[width=\linewidth]{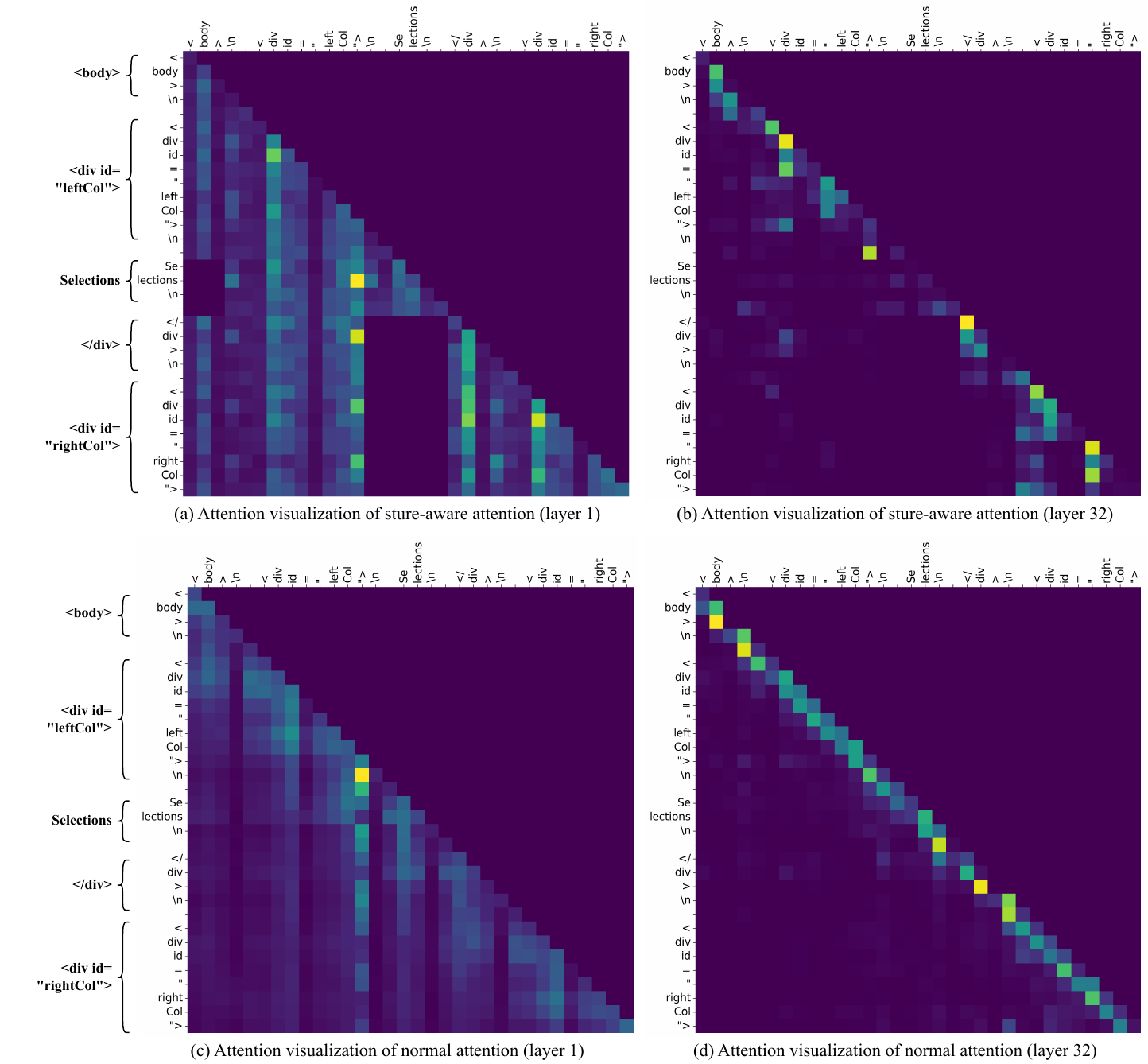}
    \caption{Selected Attention Weight Visualization of \ours{} on a simplified example.}
    \label{fig:attn_visualization}
\end{figure*}

The visualization of attention weights in Figure~\ref{fig:attn_visualization} provides insight into the behavior of structure-aware attention and standard attention mechanisms. Four representative examples are selected from different attention heads, each demonstrating distinct characteristics.

The attention in (a) exhibits a sparse and discrete distribution while effectively leveraging HTML domain knowledge. Notably, elements such as \texttt{<div id = "rightCol">} do not attend to the children of their siblings, allowing for a more efficient allocation of attention resources. This pattern suggests that structure-aware attention successfully encodes hierarchical relationships, prioritizing relevant dependencies.

The attention in (b) demonstrates a diagonalized pattern, aligning with the common attention sink phenomenon observed in deeper layers. The attention weights primarily concentrate along the diagonals, indicating that elements predominantly attend to themselves or closely positioned tokens. Despite this, the model retains structural awareness, as elements do not indiscriminately attend to siblings' children. This behavior suggests that structure-aware attention maintains domain knowledge while refining local relationships in later layers.

The attention in (c) presents a more scattered distribution, lacking clear structural constraints. The model attends to multiple elements, including those that do not directly influence the rendered appearance of the current node. A notable observation is that HTML elements tend to ignore tokens such as \texttt{\textbackslash n} and \texttt{=}, suggesting a preference for more informative tokens. 

The attention in (d) also exhibits a diagonalized pattern, indicative of attention sink behavior, where tokens predominantly attend to closely positioned elements. This pattern resembles the behavior observed in (b), yet lacks the structured constraints of structure-aware attention. The similarity between the two suggests that deeper layers across both models tend to favor local attention.

\subsection{\ours{}'s Computational Overhead}
We discuss the computational overhead of \ours{} in preprocessing, training, and inference separately.

\newsubsubsection{Preprocessing}
The only additional one-time preprocessing cost incurred for each sample and tokenizer is the construction cost of the structure-aware attention masks. The construction process for these structure-aware attention masks exhibits a complexity of $O(B \cdot L)$, where $B$ denotes the buffer length in our specialized HTML parser used for tracking HTML tag tokens, and $L$ represents the overall token length. This compares to the $O(L)$ complexity for constructing a standard causal mask. Notably, the buffer $B$ solely contains tokens of currently incomplete HTML tags, rendering its size small relative to the full token length $L$. Furthermore, this preprocessing can be executed on a CPU and is exceptionally lightweight, as it does not involve matrix multiplication operations. Once these attention masks are generated, they are reused throughout the entire training process without incurring further overhead, making this initial computational investment negligible when amortized across the full training cycle.

\newsubsubsection{Training}
The computational demands during the training phase of \ours{} are influenced by two primary components: Structure-Aware Attention and Contrastive Learning.

The introduction of Structure-Aware Attention results in \textbf{no additional computational cost} during training. This is because the only modification pertains to the values within the attention mask, while the total number of operations remains the same to that of standard attention mechanisms.

Our implementation of Contrastive Learning, the image embeddings are computed once and subsequently reused for both standard fine-tuning tasks and the contrastive learning objectives. The main additional computation arises from generating text embeddings, which requires an extra forward pass through the decoder. Other minor computational elements, such as similarity calculations and the computation of the contrastive loss, are negligible in comparison to the overall training expenditure.

In our experimental configuration, utilizing 4x NVIDIA A100 GPUs with WebSight-VLM, standard fine-tuning necessitates approximately \textbf{26 hours}. In contrast, training with \ours{} requires about \textbf{34 hours}, which constitutes a \textbf{31\% increase}. It is important to emphasize that this is a \textbf{one-time training cost} and \textbf{does not adversely affect inference speed}. We consider this increase a worthwhile investment, given the significant performance improvements achieved by \ours{}.

\newsubsubsection{Inference}
During inference, the additional computational cost introduced by \ours{} stems from parsing the generated HTML and constructing the structure-aware attention mask for newly generated tokens. This construction process is identical to the one employed during preprocessing and, similarly, remains lightweight in comparison to the model's other operational demands.

In conclusion, \ours{} introduces a {limited computational overhead during training} and imposes {minimal computation overhead during inference and preprocessing}.

\subsection{Analysis of Failure Examples}

\begin{figure*}[t]
    \centering
    \includegraphics[width=\linewidth]{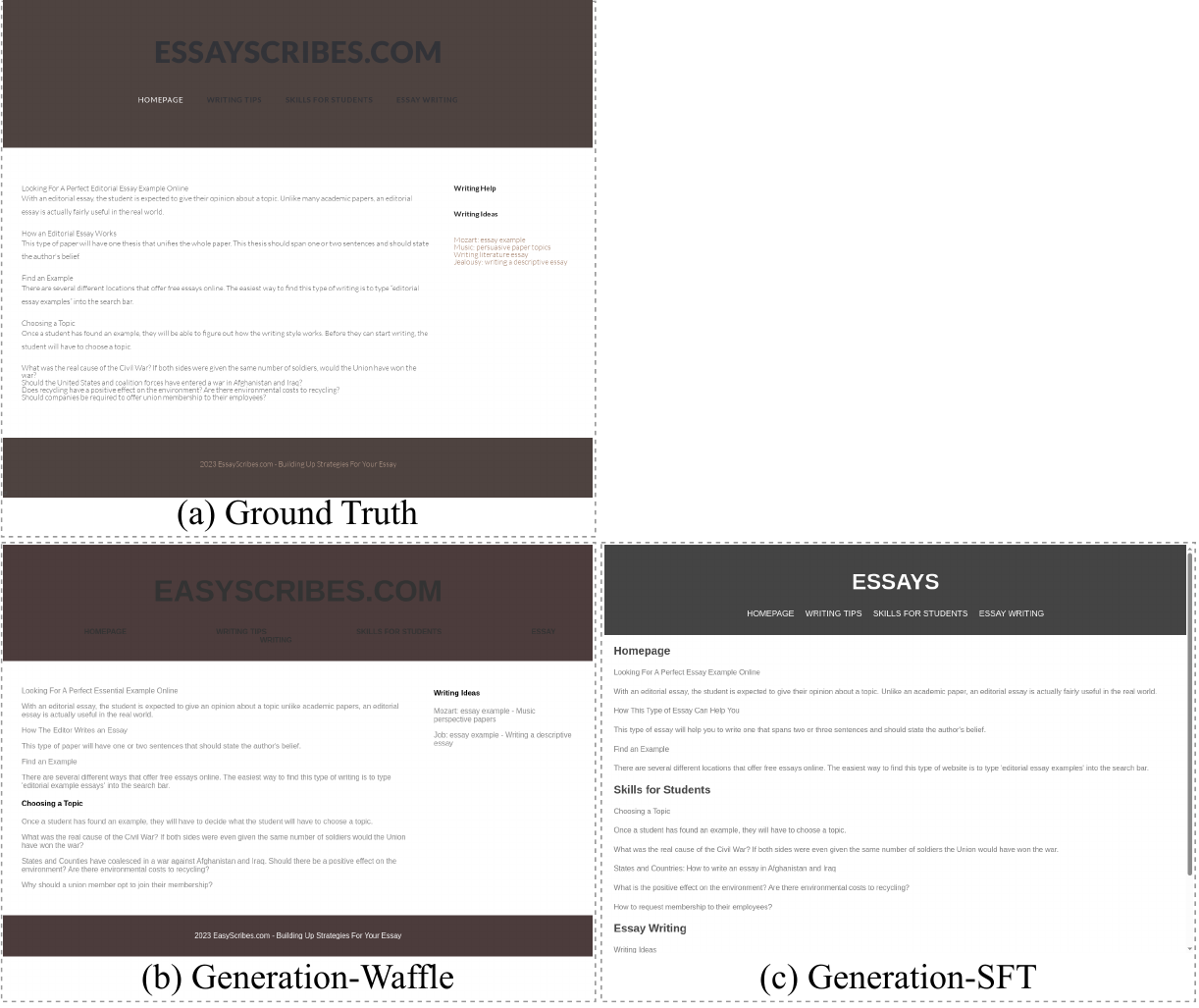}
    \caption{Figures of Design2Code task 42.}
    \label{fig:design-42}
\end{figure*}

\begin{figure*}[t]
    \centering
    \includegraphics[width=\linewidth]{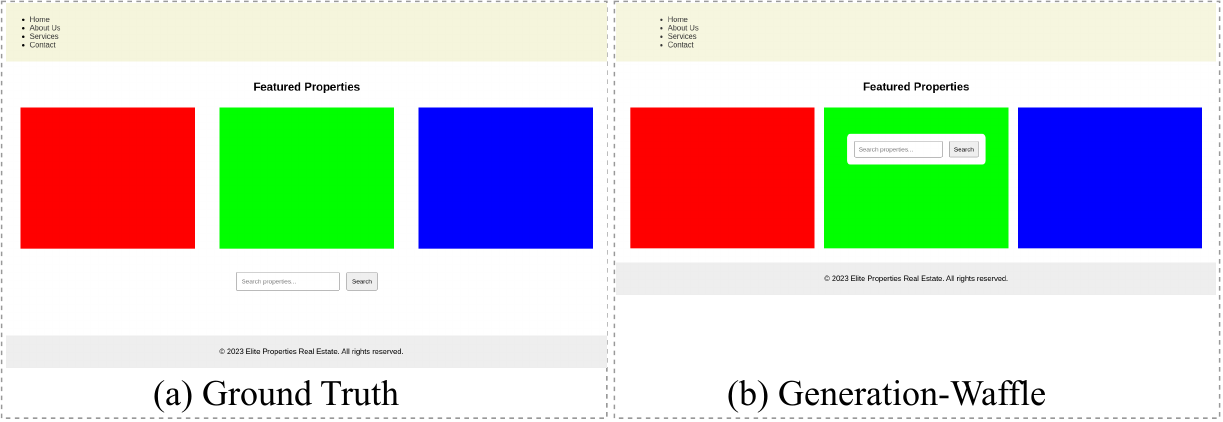}
    \caption{Figures of WebSight-Test task 45.}

    \label{fig:websight-45}
\end{figure*}

To provide a balanced evaluation of \ours{}'s capabilities, this section presents an analysis of selected failure cases. We discuss two sets of illustrative examples, one from each test set (Design2Code and WebSight-Test), where the VLM-WebSight model fine-tuned by \ours{} exhibited deviations from the ground truth.

\newsubsubsection{Design2Code-42}
The first case is from the Design2Code test set, example Design2Code-42, shown in Figure~\ref{fig:design-42}. The ground truth for this example is depicted in Figure~\ref{fig:design-42} (a), while the output generated by our \ours{}-enhanced VLM-WebSight model is shown in Figure~\ref{fig:design-42} (b). For a comparative perspective, the result from VLM-WebSight under standard fine-tuning (SFT) is provided in Figure~\ref{fig:design-42} (c).

Upon examining the output from \ours{}, several discrepancies were identified when compared to the ground truth. A primary challenge appears to be the accurate recognition and reproduction of all textual elements. For instance, there was a noticeable difference in the domain name; the ground truth specifies "ESSAYSCRIBES.COM", whereas the generated version incorrectly rendered it as "EASYSCRIBES.COM". Further textual inaccuracies included differences in the primary heading or description on the first line, where "Essential Example" was expected but variations occurred, and section headings were altered, such as "How an Editorial Essay Works" in the ground truth becoming "How The Editor Writes an Essay" in the generated HTML.

Styling inconsistencies also emerged. The navigation menu in the \ours{}-generated version exhibited different spacing and font treatment compared to the original design. Additionally, the navigation items themselves were simplified and featured slightly different labeling. Other minor errors were observed, such as in the footer area, where, although the copyright text was consistent, its formatting and color scheme diverged from the ground truth.

In summary, while VLM-WebSight fine-tuned by \ours{} successfully captured the main structure of this real-world UI from the Design2Code test set, several minor errors in text rendering and styling persist. These indicate areas for further improvement in UI-to-HTML generation. It is pertinent to note, however, that these errors are considerably less severe than those observed in the output from the VLM-WebSight model under standard fine-tuning. The SFT version exhibited more drastic failures, with significant mistakes in color fidelity, overall structure, and text extraction.

\newsubsubsection{WebSight-Test-45}
The second failure case analyzed is WebSight-Test-45. The ground truth image for this example is available as Figure~\ref{fig:websight-45} (a), and the corresponding page generated by VLM-WebSight fine-tuned with \ours{} is shown in Figure~\ref{fig:websight-45} (b).

In this instance, the generated output aligns well with the ground truth in terms of most visual elements. However, a key error was identified in the positioning of the search bar. In the generated version, the search bar is incorrectly placed at the bottom of the property cards. In contrast, the ground truth image clearly shows the search bar embedded within the middle property card. Although our \ours{}-enhanced model correctly identified that the search bar should be horizontally centered within the UI, it failed to accurately replicate its embedded placement within the designated property card.

This specific layout mistake could potentially be rectified by applying CSS rules such as \texttt{position:absolute; top:50\%; left:50\%; transform:translate(-50\%, -50\%)} to the search button's styling. The occurrence of such an error suggests that incorporating a refinement step for the generated results could be a promising avenue for future research, potentially allowing for corrections of these types of spatial misplacements.

The analysis of these failure cases offers insights into the current limitations of \ours{}, particularly in areas such as precise text recognition, complex styling fidelity, and exact component placement. These observations highlight potential directions for future improvements. 

\subsection{Infrastructure}
Our approach is implemented with the following packages:
\code{transformers 4.41.1}, \code{pytorch 2.3.0}, \code{selenium}, \code{deepspeed 0.14.1}, \code{accelerate 0.30.1}, and \code{datasets 2.19.1}. The experiments are conducted on a shared computing cluster with four NVIDIA A100 GPUs.

\subsection{Potential Risk and Impact}
This work aims to develop a multi-modality model to help front-end developers write and understand HTML and CSS code. We assume no risk of this approach being misused.

\end{document}